\documentclass[12pt]{article}
\pdfoutput=1
\usepackage{amsfonts}
\usepackage{amsmath}
\usepackage{url}
\usepackage{hyperref}
\usepackage{bbold}
\usepackage{slashed}
\usepackage{tikz}
\usepackage{graphicx}
\usepackage{epstopdf}
\usepackage{subfigure}
\usepackage{pgfplots}
\usepackage{amssymb}
\usepackage{color}
\usepackage{mathrsfs}
\usepackage{fancybox}
\usepackage{lipsum}
\usepackage{eurosym}
\usepackage{tcolorbox}

\def\({\left (}
\def\){\right )}
\def\[{\left [}
\def\]{\right ]}
\def\d{\mathrm{d}}

\numberwithin{equation}{section}

\usepackage{environ}
\NewEnviron{myequation}{%
\begin{eqnarray}
\scalebox{1.1}{$\BODY$}
\end{eqnarray}
}

\newcommand{\beq}{\begin{equation}}
\newcommand{\eeq}{\end{equation}}
 
\newcommand{\bea}{\begin{eqnarray}}
\newcommand{\ea}{\end{eqnarray}}
\newcommand{\barr}{\!\begin{array}}
\newcommand{\earr}{\end{array}\!}
\newcommand{\lb}{{\langle}}
\newcommand{\rb}{{\rangle}}

\newcommand{\threej}[6]{\left(\mbox{\small$\!\begin{array}{ccc} #1 \! & \!\! #3 \! & \!\! #5 \nspc \\[-1mm]  #2 \!  & \!\! #4 \!  & \!\! #6 \nspc \end{array}\!$}\!\right)}

\def\d{{\partial}}
\def\n{{\bf \widehat n}}
\def\k{{\bf k}}

\begin{document}
\begin{titlepage}

\setcounter{page}{1} \baselineskip=15.5pt \thispagestyle{empty}

\vfil

${}$
\vspace{1cm}

\begin{center}

\def\thefootnote{\fnsymbol{footnote}}
\begin{changemargin}{0.05cm}{0.05cm} 
\begin{center}
{\Large \bf The Schwarzian Theory - Origins}
\end{center} 
\end{changemargin}

~\\[1cm]
{Thomas G. Mertens${}^{\rm a,b}$\footnote{\href{mailto:thomas.mertens@ugent.be}{\protect\path{thomas.mertens@ugent.be}},}}
\\[0.3cm]
\vspace{0.7cm}
{\normalsize { \sl ${}^{\rm a}$Department of Physics and Astronomy,
\\[1.0mm]
Ghent University, Krijgslaan, 281-S9, 9000 Gent, Belgium}} \\
\vspace{0.5cm}
{\normalsize { \sl ${}^{\rm b}$Physics Department,
\\[1.0mm]
Princeton University, Princeton, NJ 08544, USA}} \\[3mm]

\end{center}


 \vspace{0.2cm}
\begin{changemargin}{01cm}{1cm} 
{\small  \noindent 
\begin{center} 
\textbf{Abstract}
\end{center} }
In this paper we further study the 1d Schwarzian theory, the universal low-energy limit of Sachdev-Ye-Kitaev models, using the link with 2d Liouville theory. We provide a path-integral derivation of the structural link between both theories, and study the relation between 3d gravity, 2d Jackiw-Teitelboim gravity, 2d Liouville and the 1d Schwarzian. We then generalize the Schwarzian double-scaling limit to rational models, relevant for SYK-type models with internal symmetries. We identify the holographic gauge theory as a 2d BF theory and compute correlators of the holographically dual 1d particle-on-a-group action, decomposing these into diagrammatic building blocks, in a manner very similar to the Schwarzian theory.

\end{changemargin}
 \vspace{0.3cm}
\vfil
\begin{flushleft}
\today
\end{flushleft}

\end{titlepage}

\newpage
\tableofcontents
\vspace{0.5cm}
\noindent\makebox[\linewidth]{\rule{\textwidth}{0.4pt}}
\vspace{1cm}

\addtolength{\abovedisplayskip}{.5mm}
\addtolength{\belowdisplayskip}{.5mm}

\def\plus{\raisebox{.5pt}{\tiny$+$\smpc}}

\addtolength{\parskip}{.6mm}
\def\spc{\hspace{1pt}}

\def\nspc{{\hspace{-2pt}}}
\def\ff{\rm\smpc f\smpc} 
\def\fff{\mbox{Y}}
\def\ww{{\rm w}}
\def\smpc{{\hspace{.5pt}}}

\def\zz{{\spc \rm z}}
\def\xx{{\rm x\smpc}}
\def\xxi{\mbox{\footnotesize \spc $\xi$}}
\def\jj{{\rm j}}
 \addtolength{\baselineskip}{-.1mm}

\renewcommand{\Large}{\large}

\def\calO{{b}}
\def\be{\begin{equation}}
\def\ee{\end{equation}}




\def\mathbi#1{\textbf{\em #1}} 
\def\som{{ \textit{\textbf s}}} 
\def\tom{{ \textit{\textbf t}}} 
\def\nom{n} 
\def\mom{m} 
\def\la{\langle}
\def\bea{\begin{eqnarray}}
\def\eea{\end{eqnarray}}
\def\is{\!  & \!  = \!  &  \!}
\def\ra{\rangle}
\def\half{{\textstyle{\frac 12}}}
\def\cL{{\cal L}}
\def\halfi{{\textstyle{\frac i 2}}}
\def\ba{\bea}
\def\ea{\eea}
\def\lb{\langle}
\def\rb{\rangle}
\newcommand{\rep}[1]{\mathbf{#1}}

\def\uU{\bf U}
\def\be{\bea}
\def\ee{\eea}
\def\delbar{\overline{\partial}}
\def\ra{\bigr\rangle}
\def\la{\bigl\langle}
\def\ccdot{\!\spc\cdot\!\spc}
\def\nspc{\!\spc\smpc}
\def\tr{{\rm tr}}
\def\ra{\bigr\rangle}
\def\la{\bigl\langle}
\def\li{\bigl|\spc}
\def\ri{\bigr |\spc}

\def\hf{\textstyle \frac 1 2}

\def\bfcdot{\raisebox{-1.5pt}{\bf \LARGE $\spc \cdot\spc $}}
\def\spc{\hspace{1pt}}
\def\is{\! &  \! = \! & \!}
\def\d{{\partial}}
\def\n{{\bf \widehat n}}
\def\k{{\bf k}}
\def\GO{{\cal O}}

\def\pp{{\mbox{\tiny$+$}}}
\def\mm{{\mbox{\tiny$-$}}}

\setcounter{tocdepth}{2}
\addtolength{\baselineskip}{0mm}
\addtolength{\parskip}{.4mm}
\addtolength{\abovedisplayskip}{1mm}
\addtolength{\belowdisplayskip}{1mm}

\def\fff{e}

\section{Introduction and summary}\label{sec:Sch}
\label{sect:intro}
Sachdev-Ye-Kitaev (SYK) models of $N$ Majorana fermions with random all-to-all interactions have received a host of attention in the past few years \cite{KitaevTalks, Sachdev:1992fk, Polchinski:2016xgd, Jevicki:2016bwu, Maldacena:2016hyu,Jevicki:2016ito,Turiaci:2017zwd,Gross:2017hcz,Gross:2017aos,Das:2017pif,Das:2017wae,Kitaev:2017awl}, mainly due to the appearance of maximally chaotic behavior \cite{SS, Shenker:2014cwa, MSS, Polchinski:2015cea,Turiaci:2016cvo}, suggesting a 2d holographic dual exists. It was realized immediately that the infrared behavior of these models and their relatives is given by the so-called Schwarzian theory, a 1d effective theory with action given by the Schwarzian derivative of a time reparametrization:
\begin{equation}
\label{SSch}
S_{\text{Sch}} = -C\int dt \, \left\{f,t\right\},
\end{equation}
with $\left\{f,t\right\} = \frac{f'''}{f'} - \frac{3}{2}\frac{f''^2}{f'^2}$, the Schwarzian derivative of $f$. Miraculously, the same action and interpretation appears when studying 2d Jackiw-Teitelboim (JT) dilaton gravity \cite{Jackiw:1984je, Teitelboim:1983ux,Jackiw:1992bw,Almheiri:2014cka, Jensen:2016pah, Maldacena:2016upp, Engelsoy:2016xyb, Cvetic:2016eiv,Mandal:2017thl}, with action:
\begin{equation}
\label{JTaction}
S_{JT} = \frac{1}{16\pi G_2}\int d^2x \sqrt{-g}\Phi^2\left(R^{(2)}-\Lambda\right) + S_{\text{Gibbons-Hawking}}.
\end{equation}
This leads to the holographic duality between the Schwarzian theory and Jackiw-Teitelboim gravity. UV decorations can be added to both theories if wanted, but this is the minimal theory on both sides of the duality that contains the universal gravity regime. In \cite{MTV} we solved the Schwarzian theory by embedding it in 2d Liouville CFT, fitting nicely with the well-known piece of lore that Liouville theory encodes the universal 3d gravitational features of any 2d holographic CFT. \\

A direct generalization of the SYK model is to consider instead complex fermions. These models have a U(1) internal symmetry, and the resulting infrared two-point correlator has the symmetry \cite{Davison:2016ngz}:
\begin{equation}
\label{susygreen}
G(\tau_1,\tau_2) = \left\langle \psi^{\dagger}(\tau_1)\psi(\tau_2)\right\rangle = (f'(\tau_1)f'(\tau_2))^{\Delta} \frac{g(\tau_2)}{g(\tau_1)}G(f(\tau_1),f(\tau_2)),
\end{equation}
for a function $f$, corresponding to arbitrary conformal transformations, and $g$, corresponding to arbitrary gauge transformations on the charged fermions. The former is known to be represented by a Schwarzian action, whereas the latter is represented by a free 1d particle action. \\
At large $N$ and low energies, the theory is dominated by quantum fluctuations of just these two fields. In general, the low-energy theory is then
\begin{equation}
S = -C\int dt \left(\left\{f,t\right\} + a(\partial_t g)^2\right) + S_{\text{int}}.
\end{equation}
The interaction term $S_{\text{int}}$ will depend on the specific theory at hand. Stanford and Witten \cite{Stanford:2017thb} obtained this same action by considering the coadjoint orbit action for Virasoro-Kac-Moody systems. \\
Generalizations to non-abelian global (flavor) symmetries of the fermions were studied in e.g. \cite{Yoon:2017nig,Choudhury:2017tax,Narayan:2017hvh}. \\
Finally, when considering supersymmetric SYK models with $\mathcal{N}=2$ supersymmetry, the above action (with a specific value of $a$) arises as the bosonic piece of the $\mathcal{N}=2$ super-Schwarzian action \cite{Fu:2016vas}. \\

\begin{figure}[h]
\centering
\includegraphics[width=0.95\textwidth]{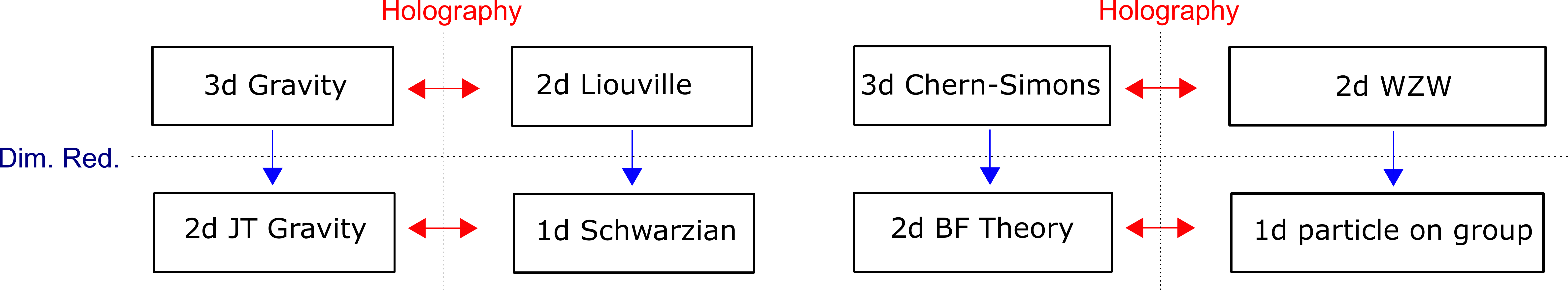}
\caption{Scheme of theories and their interrelation.}
\label{schemedimholin}
\end{figure}

Our goal here is to understand the structure behind these theories better, and their correct bulk descriptions. As a summary, we will find the following diagram of theories (Figure \ref{schemedimholin}), linking four theories through dimensional reduction and holography. The same quadrangle of theories exists for the compact group models as well.

Correlation functions of the Schwarzian theory were obtained first in \cite{altland,Bagrets:2017pwq} and generalized and put in a Liouville context in \cite{MTV}. We analogously compute correlation functions for the compact group models and find a diagram decomposition in perfect analogy with that of the Schwarzian theory in \cite{MTV}. For a compact group $G$, an arbitrary diagram is decomposed into propagators and vertices:
\bea 
\label{frules}
\begin{tikzpicture}[scale=0.57, baseline={([yshift=-0.1cm]current bounding box.center)}]
\draw[thick] (-0.2,0) arc (170:10:1.53);
\draw[fill,black] (-0.2,0.0375) circle (0.1);
\draw[fill,black] (2.8,0.0375) circle (0.1);
\draw (3.4, 0) node {\footnotesize $\tau_1$};
\draw (-0.7,0) node {\footnotesize $\tau_2$};
\draw (1.25, 1.6) node {\footnotesize $\lambda \mathbf{m}$};
\draw (6.5, 0) node {$\raisebox{6mm}{$\ \ =\ \ e^{-\spc \spc C_{\lambda} \spc (\tau_2-\tau_1)}$}$};
\end{tikzpicture}, ~~~~~~~~~~\ \ \begin{tikzpicture}[scale=0.7, baseline={([yshift=-0.1cm]current bounding box.center)}]
\draw[thick] (-.2,.9) arc (25:-25:2.2);
\draw[fill,black] (0,0) circle (0.08);
 \draw[thick](-1.5,0) -- (0,0);
\draw (.5,-1) node {\footnotesize $\textcolor{black}{\lambda_2\mathbf{m}_2}$};
\draw (.5,1) node {\footnotesize$\textcolor{black}{\lambda_1\mathbf{m}_1}$};
\draw (-1,.3) node {\footnotesize$\textcolor{black}{\Lambda \mathbf{M}}$};
\draw (3,0.1) node {$\mbox{$\ =\  \, \gamma_{\lambda_1 \mathbf{m}_1,\lambda_2 \mathbf{m}_2,\Lambda\mathbf{M}}\spc .$}$}; \end{tikzpicture}\ 
\eea
where $C_{\lambda}$ is the Casimir of the irreducible representation $\lambda$ and $\mathbf{m} \in \Omega_\lambda$ is a weight in the representation $\lambda$. The vertex function is given essentially by the $3j$-symbol of the compact group $G$:
\begin{equation}
\gamma_{\lambda_1 \mathbf{m}_1,\lambda_2 \mathbf{m}_2,\Lambda\mathbf{M}} = \threej{\lambda_1}{\mathbf{m}_1}{\lambda_2}{\mathbf{m}_2}{\Lambda}{\mathbf{M}}.
\end{equation}
The representation labels of each exterior line are summed over. In the Schwarzian theory, operator insertions are associated to discrete representations of $SL(2,\mathbb{R})$ and external lines to continuous representations, originating from the perfect dichotomy of (normalizable) states and (local) vertex operators in Liouville theory. In the rational case here, all representation labels are discrete, related to the state-operator correspondence in rational 2d CFT. \\

Our main objective is to demonstrate that the embedding of the Schwarzian theory within Liouville theory is not just convenient: it is the most natural way to think about the Schwarzian theory. This will be illustrated by both a field redefinition of Liouville theory and by immediate generalizations to compact group constructions. To expand our set of models, we also discuss $\mathcal{N}=1$ and $\mathcal{N}=2$ supersymmetric Liouville and Schwarzian theories wherever appropriate. \\

The paper is organized as follows. Section \ref{sect:path} contains a path-integral derivation of the link between Liouville theory and the Schwarzian theory. This was hinted at in \cite{MTV}, but is proven more explicitly here. We use this description of Liouville theory to exhibit more explicitly the structural links between these theories in a holographic context in section \ref{sect:class}. In section \ref{sect:BF} we look at the bulk story for the compact internal symmetries of SYK-type models. Section \ref{sect:group} discusses the 1d particle-on-a-group actions and the diagrammatic rules for computing correlation functions. We end with some concluding remarks in section \ref{sect:conc}. The appendices contain some additional technical material. \\

Recently, the papers \cite{Gonzalez:2018enk,Gaikwad:2018dfc} appeared that also investigate extensions of the Schwarzian theory with additional symmetries.

\section{Path integral derivation of Schwarzian correlators}
\label{sect:path}

In \cite{MTV} we provided a prescription for computing Schwarzian correlators through 2d Liouville theory on a cylindrical surface between two ZZ-branes. This was based on results in \cite{Dorn:2006ys,Dorn:2008sw} on (the moduli space of) classical solutions of boundary Liouville theory. Here we will provide a direct Liouville path integral derivation that substantiates our previous prescription.

\subsection{Classical limit of thermodynamics}
\label{sect:TD}
The Schwarzian limit we will take corresponds to the classical ($\hbar \to 0$) limit of a thermodynamical system.\footnote{To be distinguished from the semi-classical limit where $\beta \hbar$ is kept fixed in the limit, see e.g. \cite{Gross:1982cv}.} Let us therefore briefly review how this works. For a general theory with fields $\phi$ and momenta $\pi_\phi$, the phase space path integral of the thermal partition function is given as:
\begin{equation}
Z(\beta) = \int_{\phi(0)=\phi(\hbar\beta)} \left[\mathcal{D}\phi\right]\left[\mathcal{D}\pi_\phi\right]e^{\frac{1}{\hbar}\int_{0}^{\beta \hbar} d\tau \int dx \left(i \pi_\phi \dot{\phi} - \mathcal{H}(\phi,\pi_\phi)\right)}.
\end{equation}
Rescaling $\beta \hbar t = \tau$ and taking the classical limit, the $p \dot{q}$-term localizes to configurations with $\delta(\pi_\phi \dot{\phi}) =0$, i.e. static configurations for which $\dot{\phi} =0,\, \dot{\pi}_\phi = 0$. Hence one finds
\begin{equation}
Z(\beta) \to \int \left[\mathcal{D}\phi\right] \left[\mathcal{D}\pi_\phi\right]e^{-\beta \int dx \mathcal{H}(\phi,\pi_\phi)}
\end{equation}
which is just the classical partition function for a field configuration. \\
We will take precisely this classical limit in the Liouville phase space path integral in the next subsection.

\subsection{Gervais-Neveu field transformation}
Liouville theory with a boundary is defined by the Hamiltonian density:
\begin{equation}
\label{HamLiou}
\mathcal{H}(\phi,\pi_\phi) = \frac{1}{8\pi b^2}\left(\frac{\pi_\phi^2}{2} + \frac{\phi_\sigma^2}{2} + e^{\phi} -2\phi_{\sigma\sigma}\right).
\end{equation}
with parameters $c= 1+6Q^2$ and $Q= b+b^{-1}$. The last term integrates to a boundary term. Operator insertions in Liouville are the exponentials $V = e^{\ell \phi}$. Within the older canonical approach to Liouville theory, Gervais and Neveu \cite{Gervais:1981gs,Gervais:1982nw,Gervais:1982yf,Gervais:1983am} considered a (non-canonical) field redefinition as $(\phi,\pi_\phi) \to (A(\sigma,\tau),B(\sigma,\tau))$ with
\begin{align}
\label{GNtrans}
e^{\phi} &= -8 \frac{A_\sigma B_\sigma}{(A-B)^2}, \\
\label{GNtrans2}
\pi_\phi &= \frac{A_{\sigma\sigma}}{A_\sigma} - \frac{B_{\sigma\sigma}}{B_\sigma}-2\frac{A_\sigma+B_\sigma}{(A-B)},
\end{align}
where $A_\sigma = \partial_\sigma A$ etc. We want to apply this transformation directly in the path integral. The new functions $A$ and $B$ need to be monotonic (as can be seen from \eqref{GNtrans}): $A_\sigma \geq 0$ and $B_\sigma \leq 0$. This transformation is invertible, up to simultaneous $SL(2,\mathbb{R})$ transformations on $A$ and $B$ as:
\begin{equation}
A \to \frac{\alpha A+\beta}{\gamma A+\delta}, \quad B \to \frac{\alpha B+\beta}{\gamma B+\delta},
\end{equation}
where the quantities $\alpha$, $\beta$, $\gamma$ and $\delta$ can have arbitrary $\tau$-dependence. So we mod out by this transformation. Note that an $SL(2,\mathbb{R})$ transformation preserves the monotonicity properties of $A$ and $B$. This field redefinition \eqref{GNtrans},\eqref{GNtrans2} does not preserve the symplectic measure. \\

We are interested in the large $c$-regime (small $b$), where using this field redefinition, the Hamiltonian density \eqref{HamLiou} can be written as
\begin{align}
\label{hamil}
\mathcal{H} = -\frac{c}{24\pi}\left\{A(\sigma,\tau),\sigma\right\} -\frac{c}{24\pi} \left\{B(\sigma,\tau),\sigma\right\}.
\end{align}

The Liouville phase-space path integral, with possible insertions of the type $e^{\ell\phi}$, is then transformed into
\begin{align}
\left\langle e^{\ell\phi} \hdots \right\rangle &= \int\displaylimits_{\text{mod } SL(2,\mathbb{R})} \hspace{-1em} \left[\mathcal{D}\phi\right]\left[\mathcal{D}\pi_\phi\right] e^{\ell \phi}\hdots \, e^{\frac{c}{48\pi}\int d\sigma d\tau (i\pi_\phi \dot{\phi} - \mathcal{H})} \nonumber \\
&= \int\displaylimits_{\text{mod } SL(2,\mathbb{R})} \hspace{-1em} \left[\mathcal{D}A\right]\left[\mathcal{D}B\right] \text{Pf}(\omega)\left(\frac{A_\sigma B_\sigma}{(A-B)^2}\right)^{\ell}\hdots \, e^{\frac{c}{48\pi}\int d\sigma d\tau \left(i\pi_\phi(A,B) \dot{\phi}(A,B) +2\left\{A,\sigma\right\} + 2\left\{B,\sigma\right\}\right)}.
\label{pspi}
\end{align}
The Jacobian factor in the measure is the Pfaffian of the symplectic 2-form $\omega$. Performing the Gervais-Neveu transformation \eqref{GNtrans},\eqref{GNtrans2} on the standard symplectic measure, one finds
\begin{equation}
\label{meas}
\omega = \int_{0}^\pi d\sigma\, \delta \pi_\phi \wedge \delta \phi = \int_{0}^{\pi} d\sigma \left(\frac{\delta A''(\sigma) \wedge \delta A'(\sigma)}{A'(\sigma)^2} - \frac{\delta B''(\sigma) \wedge \delta B'(\sigma)}{B'(\sigma)^2}\right) + \text{bdy}.
\end{equation}

\begin{figure}[h]
\centering
\includegraphics[width=0.37\textwidth]{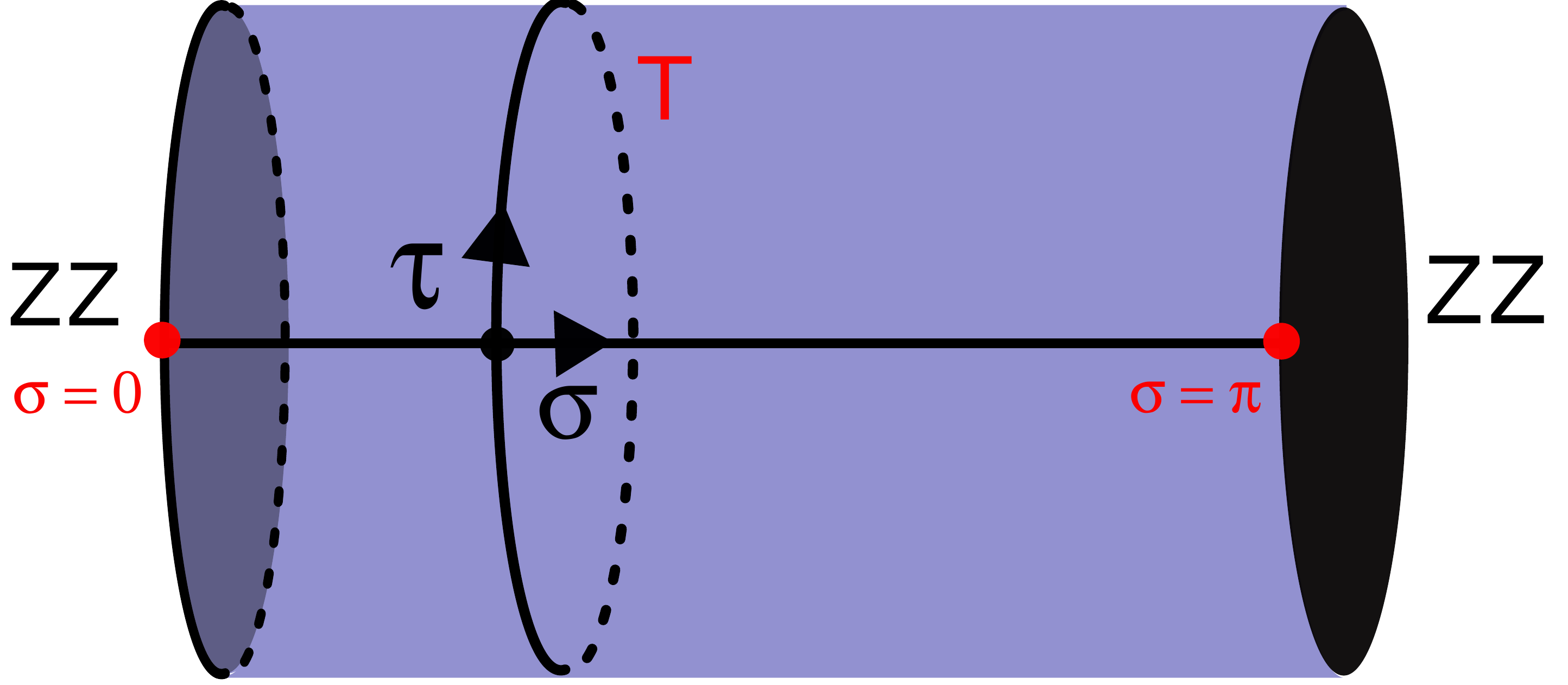}
\caption{Cylindrical surface with ZZ-branes at $\sigma=0,\pi$. The $\tau$-coordinate is chosen periodic with period $T$.}
\label{zzbrane}
\end{figure}

Next we define this theory on a cylindrical surface between two ZZ-branes \cite{Zamolodchikov:2001ah} at $\sigma =0$ and $\sigma=\pi$ (Figure \ref{zzbrane}).

The classical solution of this configuration is well-known \cite{Dorn:2006ys,Dorn:2008sw}:
\begin{equation}
e^{\phi} = -2\frac{f'(u) f'(v)}{\sin\left(\frac{f(u)-f(v)}{2}\right)^2}.
\end{equation}
in terms of a single function $f$ that satisfies $f(x+2\pi)=f(x)+2\pi$. To implement the boundary conditions at the quantum level, it is convenient to perform a thermal reparametrization of the $A$ and $B$ fields into new fields $a$ and $b$ as
\begin{equation}
\label{threp}
A(\sigma,\tau) = \tan\frac{a(\sigma,\tau)}{2}, \quad B(\sigma,\tau)= \tan\frac{b(\sigma,\tau)}{2},
\end{equation}
in terms of which \eqref{GNtrans} is rewritten as 
\begin{equation}
\label{GNtransv2}
e^{\phi} = -2 \frac{a_\sigma b_\sigma}{\sin\left(\frac{a-b}{2}\right)^2}.
\end{equation}
The redefinition \eqref{threp} preserves the monotonicity properties $a_\sigma \geq 0$ and $b_\sigma \leq 0$.
The ZZ-boundary state is characterized by $\phi\to\infty$ at the location of the branes, by \eqref{GNtransv2} requiring $\left.a=b\,\right|_{\sigma=0}$ and, by the monotonicity requirements, $\left.a=b+2\pi\,\right|_{\sigma=\pi}$. More general boundary conditions and branes are discusses in appendix \ref{app:coadjoint}. See Figure \ref{ABprofile} left. 
\begin{figure}[h]
\centering
\includegraphics[width=0.73\textwidth]{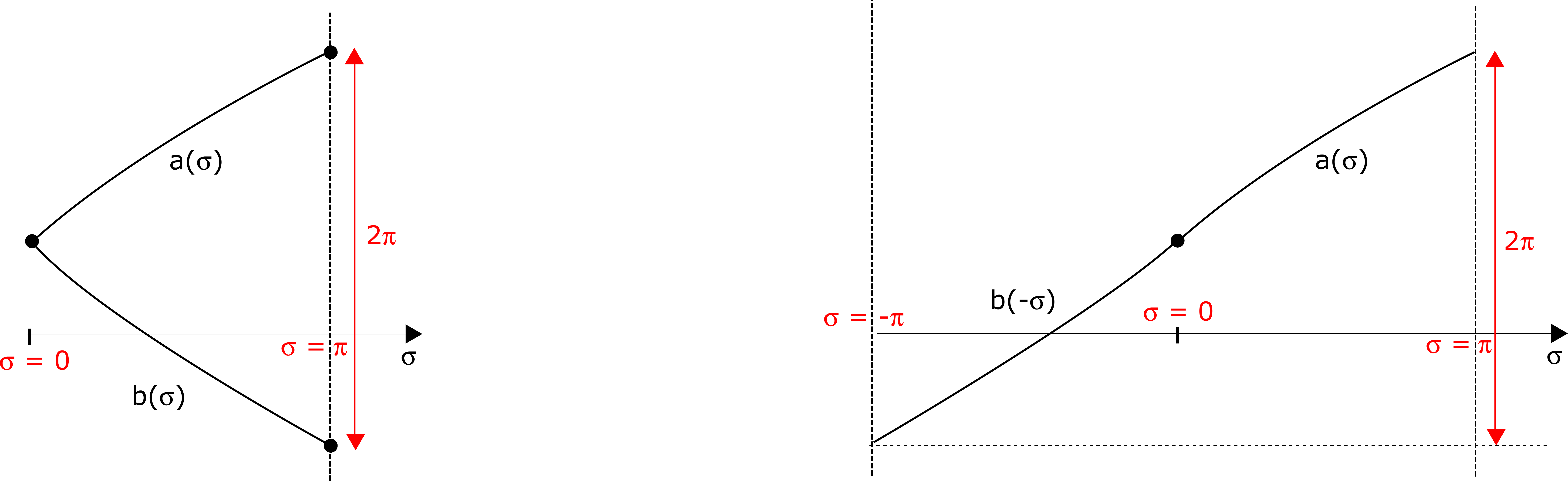}
\caption{Left: $\sigma$-dependence of $a$ and $b$ and their behavior at the branes at $\sigma=0$ and $\sigma = \pi$. Right: The doubling trick allows a description in terms of a single function $f(\sigma)$.}
\label{ABprofile}
\end{figure}

The Schwarzian limit is defined by taking the small radius limit ($T\to 0$), thereby reducing the theory to just the zero-mode along the $\tau$-direction. To obtain a theory with non-zero action, we need to take $c\to+\infty$ simultaneously such that $\frac{cT}{24\pi}=C$, a fixed constant.\footnote{Note that possible quantum renormalization effects (such as the Liouville determinant) are killed off in this limit.} This double scaling limit is identical to the classical limit of thermodynamics discussed earlier in section \ref{sect:TD}. We obtain for the Liouville correlator in this limit:\footnote{To avoid cluttering the equations, the "mod $SL(2,\mathbb{R})$" is left implicit here. In the Schwarzian limit, the arbitrary $\tau$-dependence of the $SL(2,\mathbb{R})$ transformation matrix disappears, and it becomes a global gauge redundancy.}
\begin{align}
\label{phasint}
\left\langle e^{\ell\phi} \hdots \right\rangle_{\text{ZZ-ZZ}} &= \hspace{-0.5cm}\int\displaylimits_{\tiny \left.\begin{array}{c}
a(0,\tau)=b(0,\tau) \\
a(\pi,\tau) = b(\pi,\tau)+2\pi
 \end{array}\right.} \hspace{-2em}\left[\mathcal{D}A\right]\left[\mathcal{D}B\right] \text{Pf}(\omega)\left(\frac{A_\sigma B_\sigma}{(A-B)^2}\right)^{\ell}\hdots \, e^{\frac{c}{48\pi}\int_{0}^{\pi} d\sigma \int_{0}^{T} d\tau \left(i\pi_\phi \dot{\phi} +2\left\{A,\sigma\right\} + 2\left\{B,\sigma\right\}\right)} \nonumber \\
\vspace{-3em}
\,\,&\to \int\displaylimits_{\tiny \left.\begin{array}{c}
a(0)=b(0) \\
a(\pi) = b(\pi)+2\pi
 \end{array}\right.} \hspace{-1.7em} \left[\mathcal{D}A\right]\left[\mathcal{D}B\right] \text{Pf}(\omega)\left(\frac{A_\sigma B_\sigma}{(A-B)^2}\right)^{\ell} \hdots \, e^{C \int_{0}^{\pi} d\sigma \left[\left\{A(\sigma),\sigma\right\} + \left\{B(\sigma),\sigma\right\}\right]}.
\end{align}
This can be simplified by defining a doubled field $f$ to implement the boundary conditions on the ZZ-branes, as
\begin{align}
f(\sigma) = 
\left\{
                \begin{array}{ll}
                  a(\sigma), \quad 0<\sigma<\pi, \\
                 b(-\sigma), \quad -\pi < \sigma < 0,
                \end{array}
              \right.
\end{align}
for $f$ continuous, $f_\sigma \geq 0$ everywhere, and $f$ a $1:1$ mapping from $(-\pi,\pi)$ to $(-\pi,\pi)$, so $f \in \text{diff} S^1/SL(2,\mathbb{R})$ (Figure \ref{ABprofile} right). The symplectic form \eqref{meas} in these new variables, using that $f_\sigma = a_\sigma$ and $f_\sigma = -b_\sigma$ and that both terms add up, is written as\footnote{$\beta$ should be set to $2\pi$ here. To reintroduce $\beta$ in all expressions, one places the branes at a distance $\beta/2$ and sets $A = \tan \frac{\pi}{\beta} a$ etc. Alternatively, one can redefine $C \to C \frac{2\pi}{\beta}$ and then rescale $t \to t \frac{2\pi}{\beta}$ and $f \to f \frac{2\pi}{\beta}$. This gives the field $f$ its physical dimension and demonstrates that the coupling constant $C \sim cT$ has the dimensions of length.}
\begin{align}
\label{omega}
\omega &= \int_{0}^{\pi} d\sigma \left(\frac{\delta a''(\sigma) \wedge \delta a'(\sigma)}{a'(\sigma)^2} - \left(\frac{2\pi}{\beta}\right)^2\delta a'(\sigma)\wedge \delta a(\sigma)\right) -(a \leftrightarrow b) + \text{bdy}, \nonumber \\
&= \int_{-\pi}^{\pi} d\sigma \, \left( \frac{\delta f''(\sigma) \wedge \delta f'(\sigma)}{f'(\sigma)^2} - \left(\frac{2\pi}{\beta}\right)^2\delta f'(\sigma)\wedge \delta f(\sigma)\right) + \text{bdy},
\end{align}
which is identified with the Alekseev-Shatashvili symplectic measure on the coadjoint Virasoro orbit \cite{Alekseev:1988ce,Alekseev:1990mp}. The boundary term drops out by our choice of boundary conditions, and the expression is $SL(2,\mathbb{R})$ invariant by construction. \\
The link between Liouville theory between branes and the geometric Alekseev-Shatashvili action is made in appendix \ref{app:coadjoint}. \\

Stanford and Witten showed that for a suitable choice of gauge, this becomes the standard $SL(2,\mathbb{R})$ $\prod_t 1/\dot{f}(t)$ measure \cite{Stanford:2017thb}. Regardless, the final expression for the path integral becomes
\begin{align}
\label{corrrSchw}
\int_{\text{diff} S^1/SL(2,\mathbb{R})} \left[\mathcal{D}f\right]\text{Pf}(\omega) \left(\frac{\dot{f}(t_1) \dot{f}(-t_1)}{4\sin\left(\frac{1}{2}(f(t_1)-f(-t_1))\right)^2}\right)^{\ell} \hdots e^{C \int_{-\pi}^{\pi} dt \left\{F, t\right\} }.
\end{align}
The theory is reduced to a Schwarzian system on the circle, with $F = \tan\frac{1}{2}f$. The Lagrangian $\left\{F, t\right\}$ is the analogue of \eqref{SSch} for finite temperature. In the process, Liouville operator insertions become bilocal insertions in the Schwarzian theory. Liouville stress tensor insertions are written in (\ref{hamil}) as a sum of two Schwarzian derivatives, resp. the holomorphic and antiholomorphic stress tensor. This exhausts the non-trivial Liouville operators. We end up with a Euclidean theory on the circle. \\
As stressed in \cite{MTV}, one can then extend this expression to arbitrary times for the bilocal operators to obtain the most generic Euclidean time configuration. Expressions for correlators are then obtained by taking the double scaling limit directly in the known equations in Liouville theory. 
Afterwards, one can directly Wick-rotate these to Lorentzian signature. Both of these steps are non-trivial, and the correctness of this procedure is verified by several explicit checks in \cite{MTV}. \\

To summarize, the 1d Lagrangian is the dimensional reduction of the 2d Hamiltonian, and the 2d local vertex operators become bilocal operators in the 1d theory. This is the rule we used in \cite{MTV}, and we will use this short mnemonic later on in section \ref{sect:group} when we generalize this construction beyond $SL(2,\mathbb{R})$ to arbitrary (compact) Lie groups.

\subsection{B\"acklund Transformation}
Instead of using the Gervais-Neveu parametrization \eqref{GNtrans},\eqref{GNtrans2}, we can make one more field redefinition to get a free field theory (B\"acklund transformation) by defining
\begin{align}
\phi_F &\equiv \ln \left(-A_\sigma B_\sigma \right), \\
\pi_F &\equiv \frac{A_{\sigma\sigma}}{A_\sigma} - \frac{B_{\sigma\sigma}}{B_\sigma},
\end{align}
transforming the symplectic measure again into the canonical one:
\begin{equation}
\frac{\delta A''(\sigma) \wedge \delta A'(\sigma)}{A'(\sigma)^2} - \frac{\delta B''(\sigma) \wedge \delta B'(\sigma)}{B'(\sigma)^2} = \delta \pi_F(\sigma) \wedge \delta \phi_F(\sigma),
\end{equation}
proving that the transformation $(\phi,\pi_\phi) \to (\phi_F, \pi_F)$ is canonical in field space (see e.g. \cite{Teschner:2001rv} and references therein). The Hamiltonian gets transformed into the free-field one:
\begin{equation}
H = \frac{c}{48\pi}\int_{0}^{\pi} d\sigma \left(\frac{\pi_F^2}{2} + \frac{(\partial_\sigma \phi_F)^2}{2} \right).
\end{equation}
Boundary conditions still need to be specified however, and, when written this way, the system is not suited for the doubling trick. \\

\begin{figure}[h]
\centering
\includegraphics[width=0.73\textwidth]{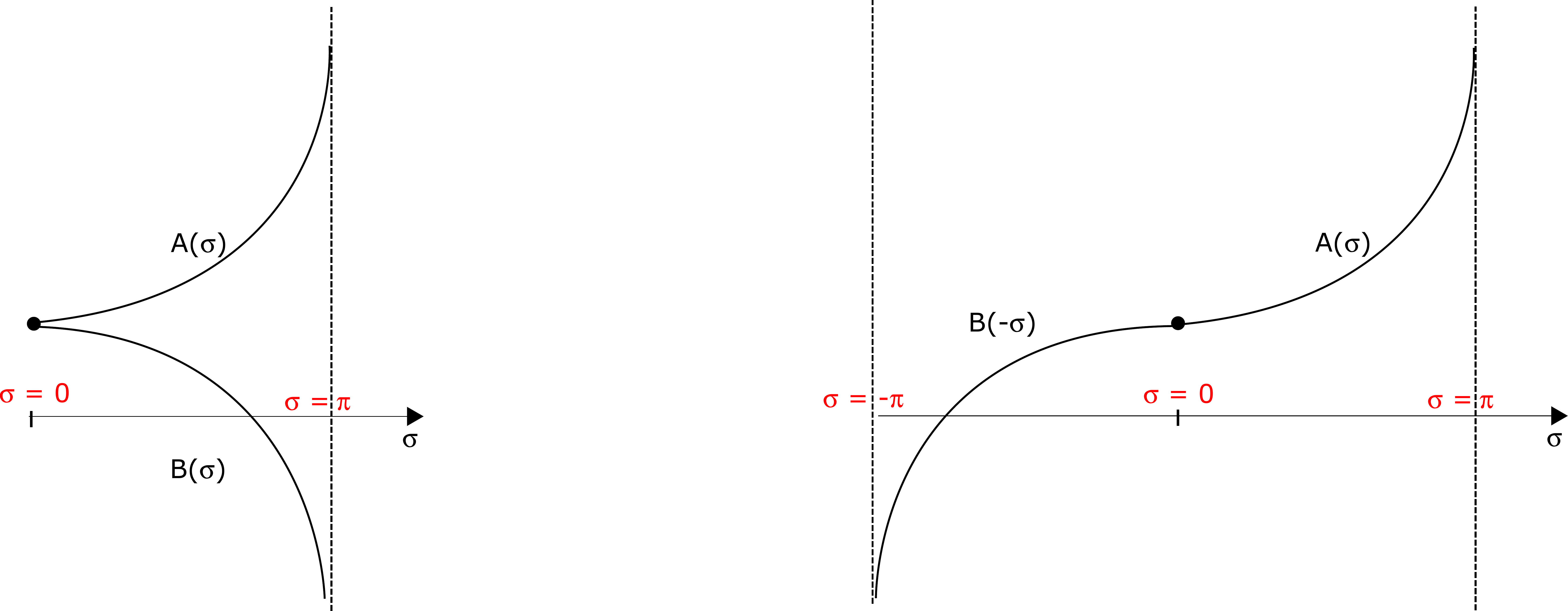}
\caption{Left: $\sigma$-dependence of $A$ and $B$ and their behavior at the branes at $\sigma=0$ and $\sigma = \pi$. Right: The doubling trick allows a description in terms of a single function $F(\sigma)$.}
\label{ABprofileb}
\end{figure}

\begin{figure}[h]
\centering
\includegraphics[width=0.75\textwidth]{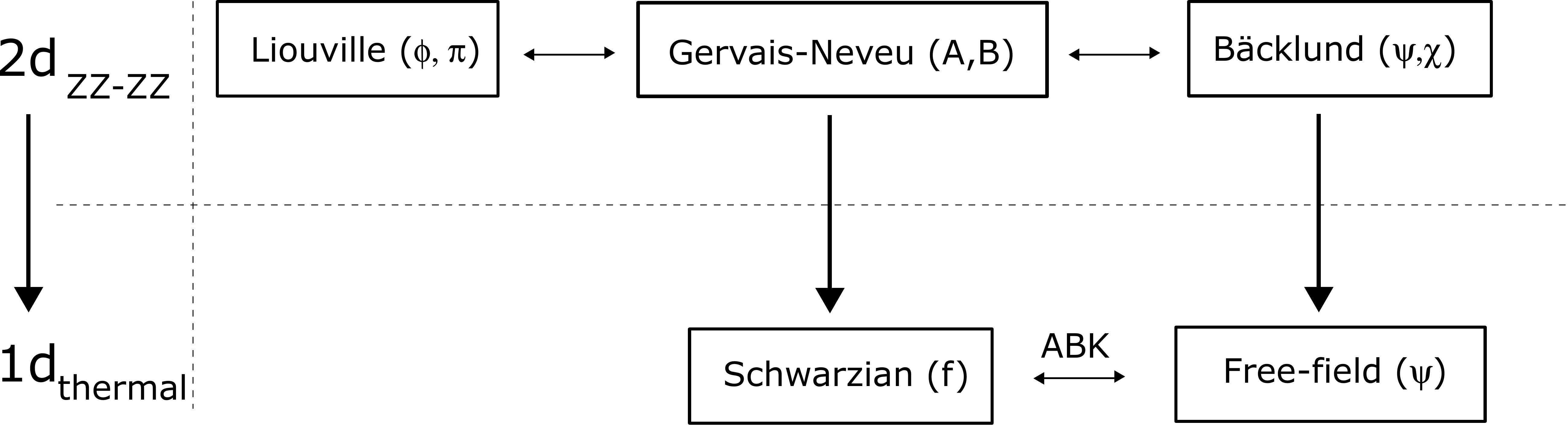}
\caption{Liouville theory in 2d in its different incarnations, and the resulting 1d theory one finds upon taking the double scaling (classical) limit. The redefinition $\dot{f}= e^{\psi}$ utilized by Altland, Bagrets and Kamenev (ABK) \cite{altland,Bagrets:2017pwq}, is the dimensional reduction of the transition from Gerveu-Neveu variables to B\"acklund variables.}
\label{schemeLiouville}
\end{figure}

There is a slight variant of this transformation that is better equiped for this purpose, by defining ($\psi,\chi$) as:
\begin{equation}
\label{ABK}
A_\sigma \equiv e^{\psi}, \quad B_\sigma \equiv -e^{\chi},
\end{equation}
or, in terms of the B\"acklund variables: $\phi_F = \psi + \chi, \quad \pi_F = \psi_\sigma -\chi_\sigma$. It will turn out that these field variables correspond to the Alekseev-Shatashvili fields \cite{Alekseev:1988ce,Alekseev:1990mp}. Upon taking the Schwarzian limit, they correspond also with the field variables utilized in \cite{altland,Bagrets:2017pwq}. \\
In these variables, $H = -\frac{c}{24\pi}\int_{0}^{\pi} d\sigma \left(\left\{A,\sigma\right\} + \left\{B, \sigma \right\}\right) = \frac{c}{48\pi} \int_{0}^{\pi} d\sigma \left((\partial_\sigma \psi)^2 + (\partial_\sigma \chi)^2\right)$. The field transformation $(\phi,\pi_\phi) \to (\psi, \chi)$ has a harmless symplectic form:
\begin{equation}
\frac{\delta A''(\sigma) \wedge \delta A'(\sigma)}{A'(\sigma)^2} - \frac{\delta B''(\sigma) \wedge \delta B'(\sigma)}{B'(\sigma)^2} = \delta \psi'(\sigma) \wedge \delta \psi(\sigma) - \delta \chi'(\sigma) \wedge \delta \chi(\sigma).
\end{equation}
The measure is now innocuous as it's field-independent, and can be readily evaluated in terms of an auxiliary fermion $\eta$ as
\begin{equation}
\text{Pf}(\omega) = \int \left[\mathcal{D}\eta\right] e^{-\int d\tau \eta' \eta} = \left(\text{det }\partial_\tau\right)^{1/2}.
\end{equation}
To implement the ZZ-boundary conditions for $\psi$ and $\chi$, we need to return to the $A$ and $B$ fields using \eqref{threp}. The boundary conditions in terms of these is illustrated in Figure \ref{ABprofileb}.
Doubling is done in terms of a single field $F$
\begin{align}
F(\sigma) = 
\left\{
                \begin{array}{ll}
                  A(\sigma), \quad 0<\sigma<\pi, \\
                 B(-\sigma), \quad -\pi < \sigma < 0,
                \end{array}
              \right.
\end{align} 
defined for the doubled interval $(-\pi,\pi)$, with $F(\pi) = F(-\pi)+\infty$, in the sense of the above figure. Defining a doubled $\psi$-field for the interval $(-\pi,\pi)$, the winding constraint is written as:
\begin{equation} 
F(\pi)=F(-\pi)+\infty \quad \to \quad \int_{-\pi}^{\pi} d\sigma \, e^{\psi} = \infty,
\end{equation}
which can be regularized and implemented in the theory using a Lagrange multiplier \cite{altland,Bagrets:2017pwq}. The path integral becomes:\footnote{The gauge symmetry implementation is more subtle now. The original invariance
\begin{equation}
F \to \frac{\alpha F + \beta}{\gamma F +\delta},
\end{equation}
is reduced to $\gamma=0$ (to fix the divergences to $\sigma=\pm \pi$ by choice) and $\beta =0$ (the transformation \eqref{ABK} undoes this redundancy). Only rescalings $F \to \alpha^2 F$ are left, which indeed correspond to shifts in $\psi$ which leave the action \eqref{abkaction} and operator insertions invariant. This leftover gauge symmetry is explicitly distilled in correlators in \cite{altland,Bagrets:2017pwq}. \\
Also, quantum renormalization effects should be taken into account when considering the 2d system as discussed in \cite{Alekseev:1988ce,Alekseev:1990mp}.}
\begin{equation}
\label{abkaction}
\int\displaylimits_{\tiny \left.\begin{array}{c}
\text{mod } SL(2,\mathbb{R}) \\
\int_{-\pi}^{\pi} d\sigma \, e^{\psi} = \infty
 \end{array}\right.} \left[\mathcal{D} \psi\right]\left(\frac{e^{\psi(\sigma_1,\tau)}e^{\psi(-\sigma_1,\tau)}}{\left(\int_{-\sigma_1}^{\sigma_1}d\sigma \, e^{\psi(\sigma,\tau)}\right)^2}\right)^{\ell} \hdots \,\, e^{\frac{c}{48\pi}\int_{-\pi}^{\pi} d\sigma \int d\tau \left(i\psi_\sigma \psi_\tau - (\partial_\sigma \psi)^2\right)}.
\end{equation}
Again taking the double scaling limit reduces this system to the expression:
\begin{equation}
\int\displaylimits_{\tiny \left.\begin{array}{c}
\text{mod } SL(2,\mathbb{R}) \\
\int_{-\pi}^{\pi} dt \, e^{\psi} = \infty
 \end{array}\right.}
\left[\mathcal{D} \psi\right]\left(\frac{e^{\psi(t_1)}e^{\psi(-t_1)}}{\left(\int_{-t_1}^{t_1}d t \, e^{\psi}\right)^2}\right)^{\ell} \hdots \,\, e^{- C \int_{-\pi}^{\pi} dt \frac{(\partial_t \psi)^2}{2}},
\end{equation}
which can be computed explicitly as shown in \cite{altland,Bagrets:2017pwq}. \\
We remark that this theory exhibits chaotic behavior, even though it looks like a free theory. Within this language, this is explicitly found in \cite{altland,Bagrets:2017pwq}, and ultimately arises due to the above constraint (introducing a 1d Liouville potential) and the non-local nature of the operator insertions. \\
These field redefinitions and their 1d Schwarzian result are summarized in Figure (\ref{schemeLiouville}).

\subsection{$\mathcal{N}=1$ super-Liouville}
The preceding discussion can be generalized to $\mathcal{N}=1$ super-Liouville theory and the $\mathcal{N}=1$ super-Schwarzian. We will be more sketchy in this paragraph, some details are left to the reader. The analogous treatment of Gervais and Neveu for $\mathcal{N}=1$ Liouville theory appeared in \cite{Arvis:1982tq,Arvis:1982kd,Babelon:1985gw} and we heavily use their results. \\
$\mathcal{N}=1$ super-Liouville theory is defined by the Hamiltonian density
\begin{equation}
\mathcal{H} = \frac{1}{\pi b^2}\left(\frac{\pi_\phi^2}{2} + \frac{\phi_\sigma^2}{2} + \frac{1}{2}e^{2\phi} - \phi_{\sigma\sigma}- i \psi_1 \psi_2 e^{\phi} + \frac{i}{2} \left(\psi_1\psi_{1\sigma} - \psi_2\psi_{2\sigma}\right)\right),
\end{equation}
for a scalar $\phi$ and two Majorana-Weyl fermions $\psi_1$ and $\psi_2$. The auxiliary field $F$ has been eliminated by its equations of motion. In superspace $(\sigma,\tau,\theta_1,\theta_2)$, the general classical super-Liouville solution for the superfield $\Phi(\sigma,\tau,\theta_1,\theta_2)$ is written as
\begin{equation}
\label{soln=1}
e^{\Phi} = \frac{(D_1 \alpha) (D_2 \beta)}{A-B-\alpha\beta},
\end{equation}
in terms of superholomorphic bosonic functions $A(x^+,\theta_1)$, $B(x^-,\theta_2)$, and their fermionic partners $\alpha(x^+,\theta_1)$, $\beta(x^-,\theta_2)$, with $D_i= \partial_{\theta_i} + \theta_i \partial_\sigma$ the superderivative. \\
As before, this can be generalized to an off-shell field redefinition in the phase space path integral:
\begin{equation}
\label{fieldred}
(\phi, \pi_\phi, \psi_1, \psi_2) \quad \to \quad (A(\sigma,\tau,\theta_1), \alpha(\sigma,\tau,\theta_1), B(\sigma,\tau,\theta_2), \beta(\sigma,\tau,\theta_2),
\end{equation}
utilizing the off-shell generalization of \eqref{soln=1} and the conjugate momentum as the definition of the non-canonical field redefinition (see \cite{Arvis:1982tq} for details). These fields are not completely independent, but satisfy 
\begin{equation}
\label{superrepa}
D_1 A = \alpha D_1 \alpha, \quad D_2 B = \beta D_2 \beta,
\end{equation}
making the transformation a super-reparametrization, and reducing the number of real components from eight to four, matching the l.h.s. of \eqref{fieldred}. In these variables, super-Liouville theory is naturally interpreted as the theory of all super-reparametrizations, generalizing this statement from previous sections. \\

To rewrite the theory in terms of these variables, consider first the differential equation 
\begin{equation}
D^3_i x = \mathcal{V}_i x,
\end{equation}
for a fermionic function $\mathcal{V}_i(\sigma,\tau,\theta_i)$. For e.g. $i=1$, one checks that this equation is solved for $x = (D\alpha)^{-1}, \, A (D\alpha)^{-1}, \, \alpha (D\alpha)^{-1}$ with $\mathcal{V}_i$ equal to (minus) the super-Schwarzian derivative, and $A$ and $\alpha$ linked by \eqref{superrepa}. Indeed, evaluating the above for e.g. $x = (D\alpha)^{-1}$ gives explicitly
\begin{equation}
\mathcal{V}_1 = D^3((D\alpha)^{-1}) D\alpha = - \frac{D^4\alpha}{D\alpha} + \frac{2D^3\alpha D^2\alpha}{(D\alpha)^2} = - \text{Sch}(\alpha,A; \sigma,\theta_1).
\end{equation}
Analogous formulas hold for $\mathcal{V}_2$ in terms of $\beta$ and $B$. \\
It was then demonstrated in \cite{Arvis:1982tq} that the Hamiltonian density can be written as
\begin{equation}
\mathcal{H} = \frac{c}{12\pi} (\mathcal{U}_1+\mathcal{U}_2),
\end{equation}
where $\mathcal{U}_i$ is the bosonic ($\sim \theta_i$) component of $\mathcal{V}_i$:
\begin{equation}
\label{Vs}
\mathcal{V}_1(\sigma,\tau,\theta_1) = \Lambda_1(\sigma,\tau) + \mathcal{U}_1(\sigma,\tau) \theta_1, \quad \mathcal{V}_2(\sigma,\tau,\theta_2) = \Lambda_2(\sigma,\tau) + \mathcal{U}_2(\sigma,\tau) \theta_2.
\end{equation}
The bosonic pieces of $\mathcal{V}_i$ thus become the Hamiltonian density in real space (after integrating over $\theta$). The fermionic parts (the $\Lambda$'s) in \eqref{Vs} are interpreted as the supercharge densities. \\

ZZ-brane boundary conditions at $\sigma =0,\pi$ require that $\Phi \to \infty$ at those locations, which means by \eqref{soln=1}, next to the bosonic conditions on $A$ and $B$, that $\left.\alpha = \pm \beta\right|_{\sigma=0,\pi}$. This again allows us to recombine $A$ and $B$ into a single reparametrization $F$, and $\alpha$ and $\beta$ into $\eta$, the superpartner of $F$. For the latter, one needs to choose $\widetilde{NS}$ (opposite) boundary conditions on the branes such that $\alpha=\beta$ on one end and $\alpha=-\beta$ on the other. This leads to an antiperiodic fermionic field $\eta$ on the doubled circle, which indeed corresponds to a thermal system. It is possible to choose other fermionic boundary conditions at the ZZ-branes, but this only leads to the $\mathcal{N}=0$ Schwarzian as discussed in \cite{MTV}. \\
Super-Liouville vertex operators $e^{\alpha \Phi}$ become bilocal super-Schwarzian operators of the form \eqref{soln=1}, given by arbitrary super-reparametrizations of the classical Liouville solution.

\section{Classical dynamics of Liouville and 3d gravity}
\label{sect:class}
Here we analyze some aspects of the classical dynamics of 2d Liouville and 3d AdS gravity with the dimensional reduction to the 1d Schwarzian and 2d Jackiw-Teitelboim gravity in mind. The larger goal is to demonstrate the structural links between 2d Liouville theory, 3d gravity, the Schwarzian theory, and JT gravity. The next section generalizes this further to other theories.

\subsection{Liouville with energy injections}
In \cite{Engelsoy:2016xyb}, we analyzed the Schwarzian theory at the classical level in 2d Jackiw-Teitelboim (JT) gravity by allowing energy injections from the boundary. We demonstrated there that the matter energy determines a preferred coordinate frame close to the boundary. Here we show how that analysis directly generalizes to the higher dimensional Liouville theory. For this purpose, the Gervais-Neveu variables ($A$, $B$) are most useful. \\
Liouville theory at large $c$ is expected to describe the universal gravitational features of holographic CFTs, and it is this regime we discuss here. 
As in \eqref{GNtrans}, the Liouville exponential is related to the ($A$, $B$) fields as
\begin{equation}
e^{\phi} \sim \frac{A_\sigma B_\sigma}{(A-B)^2}.
\end{equation}
On-shell, $A$ and $B$ are holomorphic resp. antiholomorphic functions and the Liouville metric $ds^2=e^{\phi}dx^+dx^-$ is transformed from the Poincar\'e patch into an arbitrary frame.\footnote{We take here a more general situation than in the previous section \ref{sect:path} as we do not include ZZ-branes but consider instead an infinite plane.} 

The lightcone stress tensor components are given by equation \eqref{hamil}:
\begin{equation}
\label{sstres}
T_{++}(\sigma,\tau) = -\frac{c}{24\pi}\left\{A(\sigma,\tau),\sigma\right\}, \quad T_{--}(\sigma,\tau) = -\frac{c}{24\pi}\left\{B(\sigma,\tau),\sigma\right\},\quad T_{+-}=0,
\end{equation}
leading to
\begin{equation}
T_{00}(\sigma,\tau) = T_{\sigma\sigma}(\sigma,\tau)= T_{++} + T_{--}.
\end{equation}
Energy conservation would ordinarily result in holomorphicity for $T_{++}$ and $T_{--}$. However this is violated if the system is not closed, as happens when one would inject additional energy into the system. We allow for this possibility here. 
The Schwarzian theory has its time coordinate identified with the Liouville spatial coordinate $\sigma$, thus we relabel the Liouville coordinates to reflect this: we set $\tau \to x$ and $\sigma \to t$. This corresponds to swapping the roles of time and space in Liouville theory. The total energy on a constant-$t$ slice equals
\begin{equation}
E(t) = \int d x T_{\sigma\sigma}(t,x) = \int dx \left(T_{++}(t,x) + T_{--}(t,x)\right). 
\end{equation}
Within a holographic theory with bulk coordinates ($t,r,x$), the total change in boundary energy equals the net bulk inwards flux from the boundary:
\begin{equation}
\frac{dE(t)}{dt} = -\frac{c}{24\pi} \frac{d}{dt} \int d x \left(\left\{A(t,x),t\right\} + \left\{B(t,x),t\right\}\right) = -\int dx \, T_{0r}(t,x,r\to +\infty).
\end{equation}
This equation is not that powerful in general. However, when reducing to the spatial ($=x$) zero-mode, it becomes the classical Schwarzian equation of motion \cite{Jensen:2016pah, Maldacena:2016upp, Engelsoy:2016xyb}; the Schwarzian equation is just energy conservation. \\
When evaluating \eqref{sstres} on a region where energy is conserved, all functions become holomorphic and this just reduces to the uniformizing coordinate identification:
\begin{equation}
T_{++}(x^+) = -\frac{c}{24\pi}\left\{A(x^+),x^+\right\}, \quad T_{--}(x^-) = -\frac{c}{24\pi}\left\{B(x^-),x^-\right\},
\end{equation}
where $x^\pm = \tau \pm \sigma$. 

\subsubsection{Bulk interpretation}
\begin{figure}[h]
\centering
\includegraphics[width=0.50\textwidth]{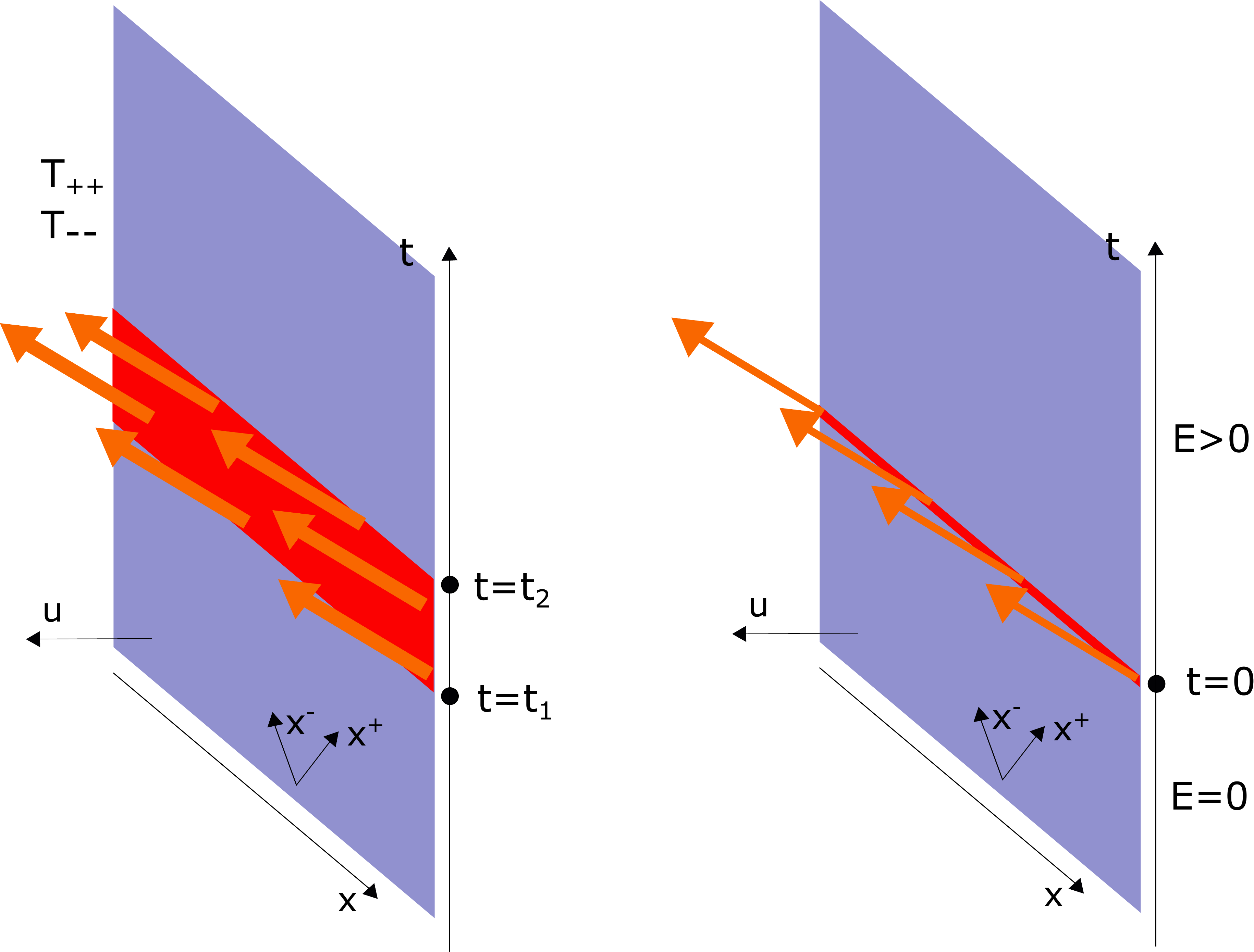}
\caption{Left: classical injection of bulk energy between $t_1 < t < t_2$. We consider the region after the injection takes place $t>t_2$ where a non-zero boundary $T_{\pm\pm}$ was generated. Right: classical injection of a translationally symmetric pulse into the bulk.}
\label{liouvilleclassical}
\end{figure}

The above can be interpreted as a diffeomorphism from vacuum Poincar\'e AdS$_3$ ($A,B$) into a new preferred frame ($x^+,x^-$). It is clearest to demonstrate this in a region where no additional matter falls in (or is extracted) (Figure \ref{liouvilleclassical} left).
It has been shown in \cite{Roberts:2012aq} that the general bulk diffeomorphism that brings the Poincar\'e AdS$_3$ solution $(X^+,X^-,u)$
\begin{equation}
ds^2 = \frac{-2dX^+dX^- + du^2}{u^2},
\end{equation}
to the Banados metric $(x^+,x^-,z)$
\begin{equation}
ds^2 = L_+(x^+) dx_+^2 + L_-(x^-) dx_-^2 + \left(-\frac{2}{z^2} + \frac{z^2}{2}L_+(x^+)L_-(x^-)\right)dx_+ dx_- + \frac{dz^2}{z^2},
\end{equation}
is found by extending the transformation
\begin{align}
\label{bulkdiffcl}
X^\pm = X^\pm(x^\pm) + \mathcal{O}(z^2), \quad u = z\sqrt{\partial X^+(x^+) \partial X^-(x^-)} + \mathcal{O}(z^3)
\end{align}
into the bulk, with the chiral functions $X^\pm(x^\pm)$ and $L_\pm(x^\pm)$ determined by solving\footnote{
The full bulk diffeomorphism is given by
\begin{align}
\label{bulkdiffeo}
X^{\pm} = X^{\pm}(x^\pm) + \frac{2z^2 {\partial_{\pm}{X}^{\pm}}^2 \partial_{\mp}^2 X^{\mp}}{8\partial_{+} X^{+} \partial_{-} X^- - z^2 \partial_{+}^2 X^+ \partial_{-}^2 X^-}, \quad\quad
u = z \frac{\left(4\partial_{+} X^{+} \partial_{-} X^-\right)^{3/2}}{8\partial_{+} X^{+} \partial_{-} X^- - z^2 \partial_{+}^2 X^+ \partial_{-}^2 X^-}.
\end{align}

}
\begin{equation}
\label{banastress}
-\frac{1}{2}\left\{X^\pm,x^\pm\right\} \equiv L_{\pm}(x^\pm) = \frac{12\pi}{c} T_{\pm\pm}(x^\pm),
\end{equation}
given $T_{\pm\pm}$. Then $X^+(x^+) = A(x^+)$ and $X^-(x^-) = B(x^-)$. Hence the functions $A$ and $B$ indeed correspond to the boundary reparametrization that, upon extending into the bulk using \eqref{bulkdiffeo}, is precisely the required frame. Setting $z=\epsilon$ in \eqref{bulkdiffcl} leads to a radial trajectory $u(X^+,X^-)$ representing a fluctuating holographic boundary caused by matter injections. Note that solving \eqref{banastress} directly leaves a $SL(2,\mathbb{R}) \times SL(2,\mathbb{R})$ ambiguity, which is fixed by boundary (gluing) conditions, just as in the 2d case \cite{Engelsoy:2016xyb}. \\

As an explicit example, consider a translationally invariant injection of matter through a pulse (Figure \ref{liouvilleclassical} right).
This requires $T_{++} = T_{--}$ to set $T_{tx}=0$ for $t>0$, equal to (half) the energy injected. One can then immediately solve \eqref{banastress} for $A$ and $B$ after the pulse:
\begin{align}A(x^+)&= \left\{
                \begin{array}{ll}
                  x^+, \quad t<0\\
                  \sqrt{\frac{c}{12 \pi E}}\tanh\left(\sqrt{\frac{12 \pi E}{c}} x^+\right), \quad t>0,
                \end{array}
              \right. \\
B(x^-)&= \left\{
                \begin{array}{ll}
                  x^-, \quad t<0\\
                  \sqrt{\frac{c}{12 \pi E}}\tanh\left(\sqrt{\frac{12 \pi E}{c}} x^-\right), \quad t>0.
                \end{array}
              \right.
\end{align}
The resulting Banados metric at $t>0$ is of course the BTZ black hole frame.\footnote{Note that these functions are not strictly holomorphic, due to the jump at $t=0$. This was indeed allowed in regions where energy is not conserved.}

\subsection{Jackiw-Teitelboim from 3d}
\label{sect:bulkschwlim}
It has been known for a long time that a spherical dimensional reduction of 3d gravity yields 2d Jackiw-Teitelboim gravity \cite{Achucarro:1993fd}. This is done by considering the 3d ansatz
\begin{equation}
ds^2 = g^{(2)}_{\mu\nu}dx^\mu dx^\nu + \lambda^{-2} \Phi^4 d\varphi^2,
\end{equation}
with $\lambda$ a mass scale. This yields directly\footnote{$\Lambda = -\frac{2}{L^2}$.}
\begin{equation}
\frac{1}{16\pi G_3} \int d^3x \sqrt{-G}\left(R^{(3)}-\Lambda\right) = \frac{2\pi}{16\pi \lambda G_3} \int d^2x \sqrt{-g}\Phi^2\left(R^{(2)}-\Lambda\right),
\end{equation}
which is indeed JT gravity. \\
The Schwarzian coupling constant $C \sim 1/G_2$, but $\frac{G_3}{L} \to 0$ to match 3d gravity with 2d Liouville theory at large central charge, with Brown-Henneaux central charge $c=\frac{3L}{2G_3}$. So we choose $\lambda L \to + \infty$ to obtain a finite limit with $G_2 \sim \lambda G_3$. This is the Schwarzian double scaling limit from the bulk perspective. \\
This 3d perspective on the bulk is very useful, and we here mention some aspects that become easier to understand when embedding the theory in 3d. 

\subsubsection{Black hole solutions from 3d}
At the level of classical solutions, the general vacuum solution of 3d $\Lambda <0$ gravity is the Banados metric:
\begin{equation}
ds^2 = L_+ dx_+^2 + L_- dx_-^2 + \left(-\frac{2}{z^2} + \frac{z^2}{2}L_+L_-\right)dx_+ dx_- + \frac{dz^2}{z^2},
\end{equation}
for arbitrary chiral functions $L_\pm(x^\pm)$. \\
Performing a spherical dimension reduction requires $L_+=L_- = L$, a constant, as it should be independent of $\varphi$. The resulting 3d space is a non-rotating BTZ black hole, dimensionally reducing to a 2d JT black hole. \\
By \eqref{banastress}, only constant Schwarzian solutions survive the reduction, as this is the generic 3d metric outside matter. And any 2d vacuum metric in JT theory is a black hole of a given mass. Indeed, directly solving the vacuum JT equations (as in \cite{Almheiri:2014cka, Jensen:2016pah, Maldacena:2016upp, Engelsoy:2016xyb, Cvetic:2016eiv}) leads to black hole spacetimes as the only solutions, perfectly analogous to the 2d CGHS models \cite{Callan:1992rs}.

\subsubsection{Fefferman-Graham from 3d}
In \cite{Almheiri:2014cka, Jensen:2016pah, Maldacena:2016upp, Engelsoy:2016xyb}, JT gravity is defined by enforcing an asymptotic value $\Phi^2 \sim a/\epsilon$ of the dilaton $\Phi^2$ at $z=\epsilon$, combined with an asymptotically Poincar\'e metric. Here we demonstrate that, upon embedding in 3d, both of these conditions follow from just imposing asymptotically Poincar\'e boundary conditions directly in 3d.
The 3d BTZ metric can be written as
\begin{equation}
ds^2 = -4(\rho^2-\mu a)\frac{dt^2}{a^2} + \frac{d\rho^2}{\rho^2-\mu a} + \rho^2 d\varphi^2.
\end{equation}
Performing the purely radial transformation $\rho = \sqrt{\mu a}\text{coth}\left(\sqrt{\frac{\mu}{a}}(x^+-x^-)\right)$ \cite{Almheiri:2014cka, Engelsoy:2016xyb}, and setting $ t= \frac{x^+ + x^-}{2}$, the metric becomes
\begin{equation}
ds^2 = -4 \frac{\mu dx^+ dx^-}{a \sinh\left(\sqrt{\frac{\mu}{a}}(x^+-x^-)\right)^2} + \mu a\text{coth}\left(\sqrt{\frac{\mu}{a}}(x^+-x^-)\right)^2 d\varphi^2,
\end{equation}
which is of the form of a spherical dimensional reduction:
\begin{equation}
ds^2 = g_{\mu\nu}dx^\mu dx^\nu = h_{ij} dx^i dx^j + \Phi^4(x) d\varphi^2,
\end{equation}
giving the 2d JT black hole metric $h_{ij}$ and associated dilaton field $\Phi^2$. Asymptotically, the above 3d metric behaves as
\begin{equation}
ds^2 \approx -\frac{dx^+ dx^-}{z^2} + \frac{a^2 d\varphi^2}{z^2},
\end{equation}
which, upon absorbing $a$ in $\varphi$, is just the standard Fefferman-Graham asymptotic expansion. Hence imposing Fefferman-Graham gauge in 2d and $\Phi^2 \sim a/\epsilon$ is equivalent to imposing Fefferman-Graham gauge in 3d. 

\subsection{3d embedding}
\begin{figure}[h]
\centering
\includegraphics[width=0.65\textwidth]{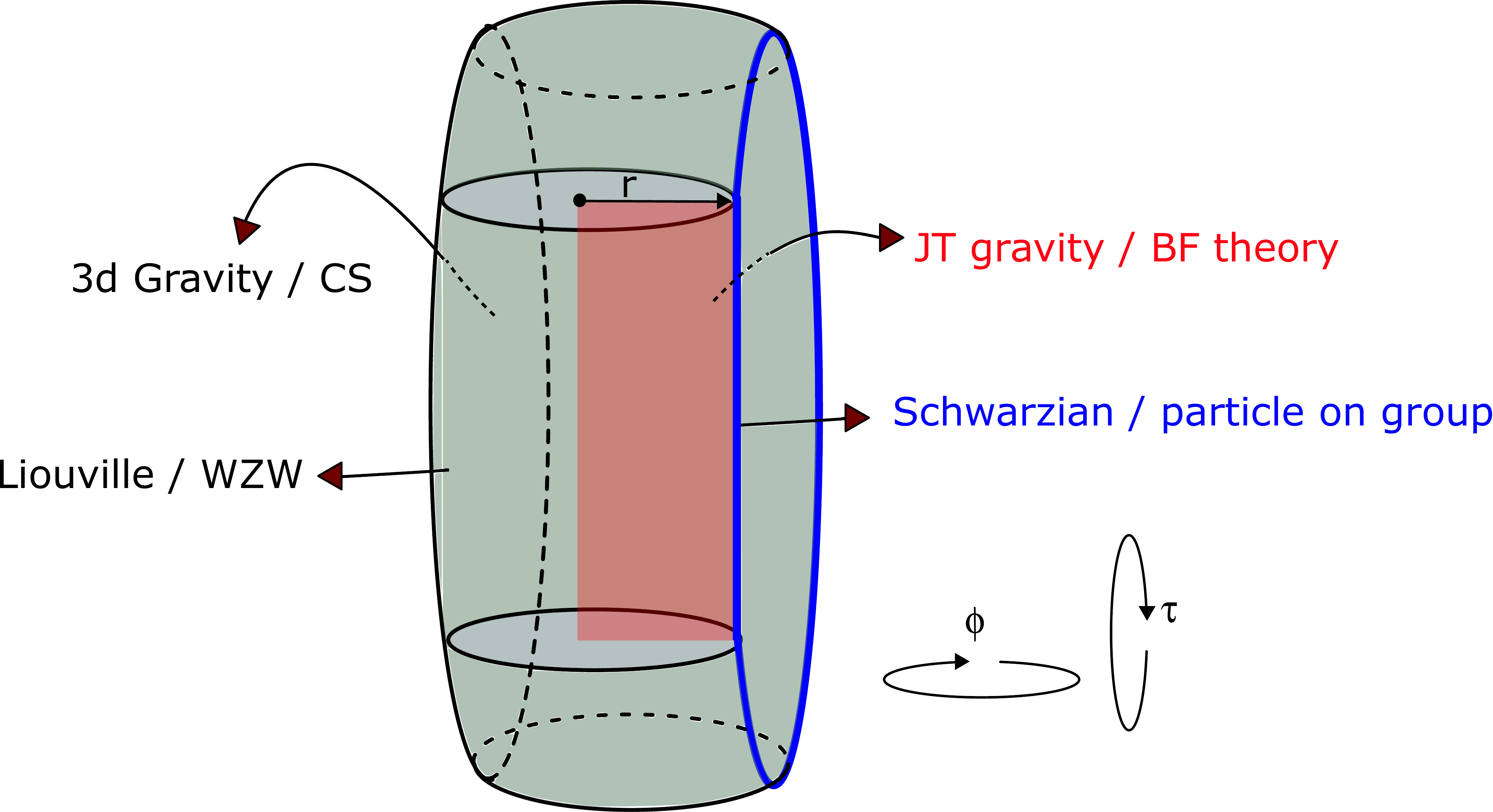}
\caption{Link between four theories through dimensional reduction, both for the gravity sector, as for the group theory sector. The interior of the torus is the 3d bulk. The torus itself is the holographic boundary. Reducing to the angular zero-mode gives a 2d bulk and a 1d boundary line.}
\label{schemedimred}
\end{figure}

Armed with the above embedding of the Schwarzian theory within Liouville and JT gravity within 3d gravity, we can now relate four different theories through dimensional reduction and the Schwarzian limit. \\
One starts with 3d gravity in the bulk, with periodically identified Euclidean time $\tau$. Its boundary contains 2d Liouville theory. Instead reducing to the angular $\varphi$-zero-mode, one obtains 2d JT gravity in the bulk. These two 2d theories are living in distinct regions and are only linked through this higher-dimensional story. Finally dimensionally reducing Liouville theory leads to the Schwarzian theory as the angular zero-mode of the boundary theory (Figure \ref{schemedimred}). 

\begin{figure}[h]
\centering
\includegraphics[width=0.95\textwidth]{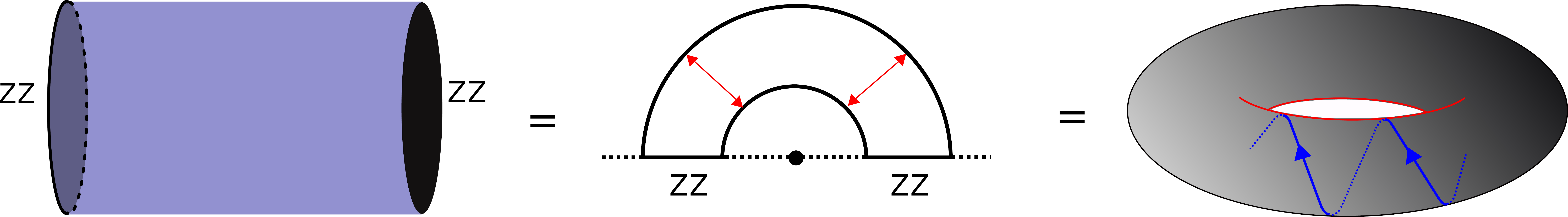}
\caption{Left: cylindrical surface bounded by ZZ-branes. Middle: The exponential map transforms this into an annular region in the upper half plane. The ZZ-branes are on the real axis and the semicircles are identified as shown. Right: Performing the doubling trick (method of images) leads to a torus with only one chirality.}
\label{chiraltorus}
\end{figure}

We can omit the ZZ-branes if we realize that their entire goal in life is to combine left- and right moving sectors into one periodic field, thereby transforming the cylindrical surface into a (chiral) torus. This equivalence is also demonstrated in Figure \ref{chiraltorus}. The propagation of just the identity module along the smaller circle is a consequence of taking the large $c$ limit.

\begin{figure}[h!]
\centering
\includegraphics[width=0.95\textwidth]{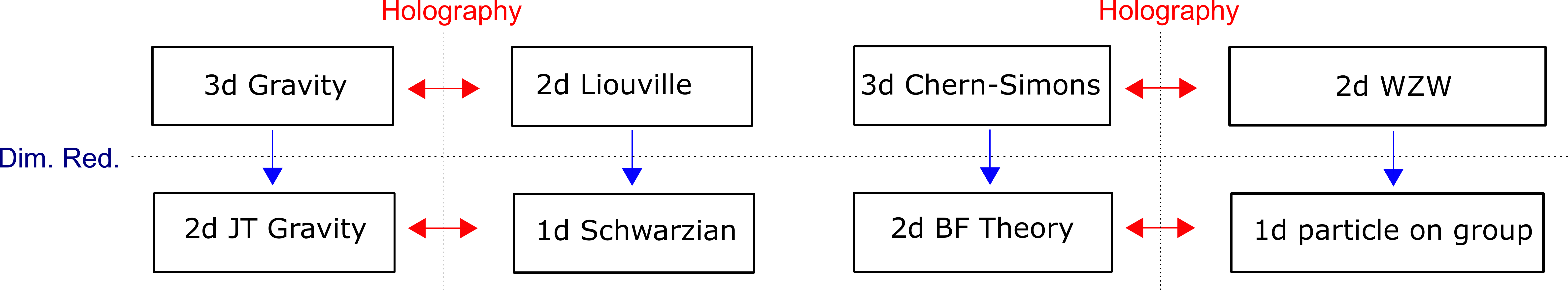}
\caption{Scheme of theories and their interrelation.}
\label{schemedimhol}
\end{figure}

As we will demonstrate starting from the next section, an analogous story holds for group theory: Chern-Simons (CS) in 3d reduces to 2d WZW on the boundary. Instead restricting to the angular zero-mode leads to 2d BF theory in a different region. Further dimensionally reducing the boundary theory leads to the 1d particle on a group manifold. The resulting scheme of models was already shown in Figure \ref{schemedimholin} and is repeated in Figure \ref{schemedimhol} for convenience.

\section{2d BF theory}
\label{sect:BF}
\subsection{Bulk derivation}
\begin{figure}[h]
\centering
\includegraphics[width=0.45\textwidth]{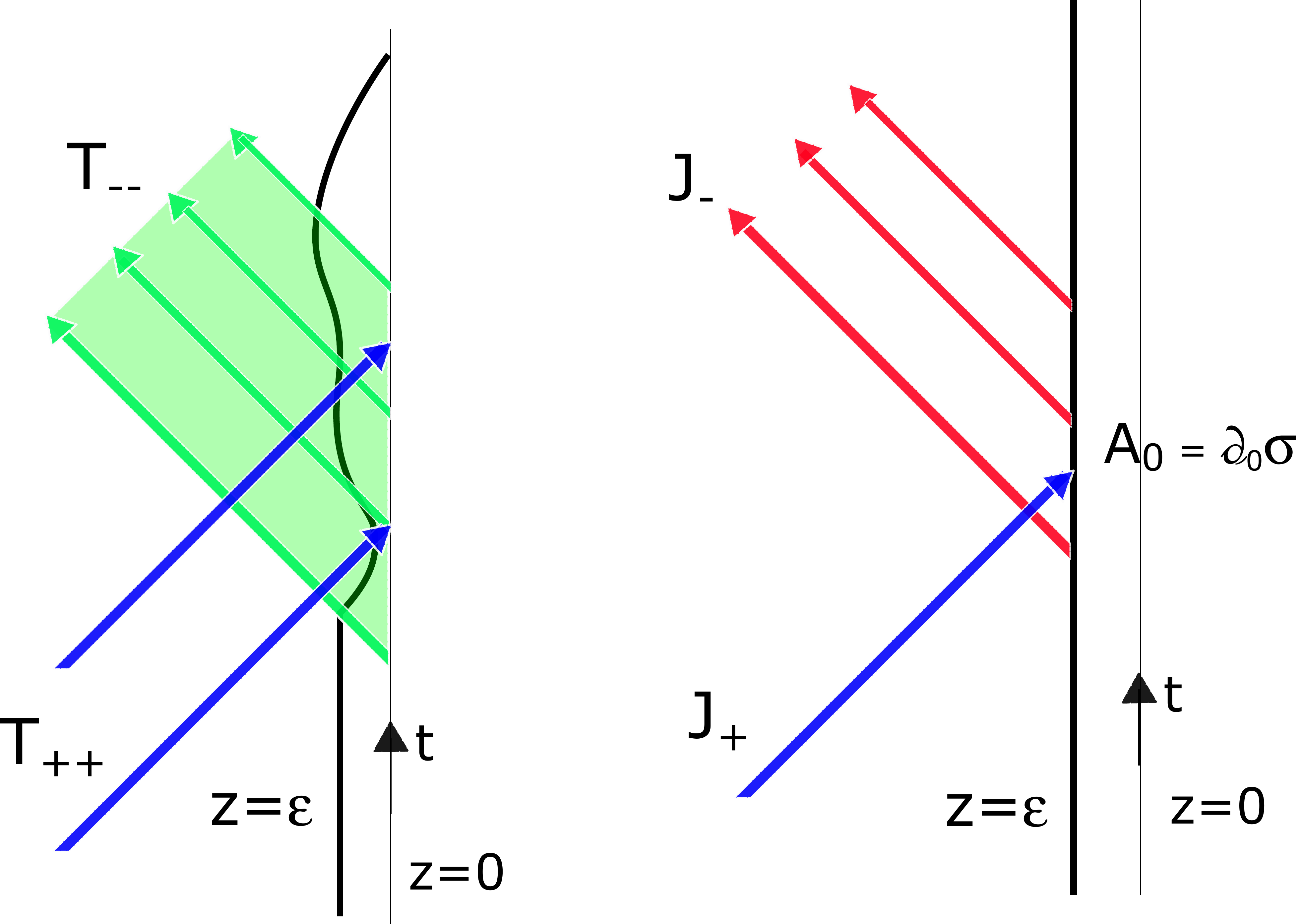}
\caption{Left: injecting energy in JT gravity leads to a preferred coordinate frame at each time, resulting in a fluctuating boundary line. Right: injecting charge leads to a preferred gauge transformation at each time.}
\label{wiggly}
\end{figure}

It was suggested in \cite{Jensen:2016pah, Maldacena:2016upp, Engelsoy:2016xyb} that the Schwarzian theory is holographically dual to Jackiw-Teitelboim gravity. Within JT gravity, the Schwarzian appears as follows. The dilaton field blows up near the AdS boundary, with a coefficient depending on the matter sector. Keeping fixed its asymptotics, requires performing a coordinate transformation at each instant, depending on the injected / extracted energy from the system. This results in a fluctuating boundary curve (Figure \ref{wiggly} left).
One can directly deduce the Schwarzian action from the bulk 2d JT dilaton gravity theory from the Gibbons-Hawking boundary term \cite{Maldacena:2016upp}. This argument has been generalized to $\mathcal{N}=1$ and $\mathcal{N}=2$ JT supergravity in \cite{Forste:2017kwy} and \cite{Forste:2017apw} respectively. In appendix \ref{app:mattersch} we extend the argument (in the bosonic case) to include an arbitrary matter sector. \\

The gauge theory variant of this story is readily formulated: we need a preferred gauge transformation on the boundary curve at each instant, determined by the injected charge into the system (see Figure \ref{wiggly} right). The correct bulk theory that describes this situation is 2d BF theory. \\
The argument we present is a dimensional reduction of the 3d Chern-Simons story and the direct analog of the Schwarzian argument of \cite{Maldacena:2016upp}. Consider the 2d BF theory obtained as a dimensional reduction from 3d CS theory:
\begin{equation}
S_{CS} \sim \int_{M_3}d^3x \epsilon^{ijk}A_i \partial_j A_k,
\end{equation}
with $A_\phi \sim \chi$ and $\partial_\phi = 0$. One obtains:\footnote{Reintroducing the correct prefactor $\frac{k}{4\pi}$ in the Chern-Simons action, by analogy with section \ref{sect:bulkschwlim}, one needs to set $A_\phi \sim \frac{\chi}{k}$ to find a finite limit. The resulting 2d action is proportional to some $C$ again, which is not quantized even though the original $k$ is.}
\begin{equation}
\label{act2d}
S = \int_M d^2x \chi F + \frac{1}{2}\oint_{\partial M} dt \chi A_0.
\end{equation}
This action is not gauge-invariant, but changes as
\begin{equation}
\delta_g S = \frac{1}{2}\oint_{\partial M} dt \chi \delta_g A_0,
\end{equation}
just like 3d CS theory. Restricting the gauge transformations to satisfy $\left.\delta_g A_0 = 0\right|_{\partial M}$, solves this problem, but creates dynamical degrees of freedom at the boundary. \\
Sending in charge through a matter field requires the additional term
\begin{equation}
S_{\text{matter}} = \int_{M} d^2x A_\mu J^\mu,
\end{equation}
which is the charge analogue of the energy-momentum matter source for the gravitational field given in appendix \ref{app:mattersch}. Varying w.r.t. $A_\mu$ and $\chi$ gives the equations of motion:
\begin{equation}
\label{BFeom}
F = 0, \quad \nabla_\mu \chi = \epsilon_{\mu\nu}J^\nu,
\end{equation}
and the boundary terms at $r=+\infty$:
\begin{equation}
\frac{1}{2}\oint_{\partial M} \left(A_0 \delta \chi - \chi \delta A_0\right).
\end{equation}
These can be cancelled by constraining:
\begin{equation}
\label{bdy}
\left.v \, \chi = A_0 \right|_{\partial M},
\end{equation}
for a parameter $v$ that defines the specific theory. We choose $v=1$. \\
Path integrating \eqref{act2d} over $\chi$ sets $F=0$ in the bulk. So we parametrize the solution as
\begin{equation}
A_\mu = \partial_\mu \sigma.
\end{equation}
Using the boundary condition \eqref{bdy}, the full action \eqref{act2d} now becomes:
\begin{equation}
S = \frac{1}{2}\oint_{\partial M} dt \, \dot{\sigma}^2.
\end{equation}
The total boundary charge is defined as
\begin{equation}
Q = \frac{\delta S_{\text{on-shell}}}{\delta A_{0}} = \dot{\sigma},
\end{equation}
and the total boundary energy is 
\begin{equation}
\label{csene}
T_{tt} = \frac{\dot{\sigma}^2}{2}.
\end{equation}
For the matter action $S_{\text{matter}}$, after integrating by parts, one finds the boundary term:
\begin{equation}
S_{\text{matter}} = -\oint_{\partial M} dt\, \sigma\, J^r = -\oint_{\partial M} dt\, \sigma\, (J_+ - J_-),
\end{equation}
representing the net inward flux of charge. \\
As charge is sent in, one requires $A_0$ to change as well asymptotically to keep fixed the boundary condition \eqref{bdy}. Either by using $\chi = \dot{\sigma}$ and \eqref{BFeom}, or by directly varying the boundary action in terms of $\sigma$, one obtains
\begin{equation}
\ddot{\sigma}= J^{r},
\end{equation}
which determines how the gauge transformation $\sigma$ evolves due to matter charge; $\sigma$ was pure gauge in the bulk but becomes physical on the boundary. \\

Some Comments:
\begin{itemize}
\item This procedure is independent of the gravity (Schwarzian) part. $\mathcal{N}=2$ JT supergravity would fix the relative coefficient (see section \ref{sect:susy} below).
\item Non-abelian generalization is straightforward. The non-abelian BF theory is
\begin{equation}
S = \int_M d^2x \text{Tr}\chi F + \frac{1}{2}\oint_{\partial M} dt \text{Tr}\chi A_0,
\end{equation}
which is gauge-invariant ($\chi$ transforms in the adjoint representation), up to the boundary term again. The equations of motion require $A_\mu = g^{-1}\partial_\mu g$, with $F=0$. The boundary condition is again chosen as $\left.\chi = A_0 \right|_{\partial M}$. So the full theory reduces to the boundary action:
\begin{equation}
S = \frac{1}{2}\oint_{\partial M} dt \text{Tr}(g^{-1}\partial_t g)^2,
\end{equation}
which is the action of a particle on a group manifold, to be studied more extensively in section \ref{sect:group} below.
\item One can write Jackiw-Teitelboim itself as an $SL(2,\mathbb{R})$ BF theory \cite{Jackiw:1992bw}, see also \cite{Grumiller:2016dbn} for recent developments. In fact, dimensionally reducing $SL(2,\mathbb{R})$ CS theory just gives us the $SL(2,\mathbb{R})$ BF theory, which is the first-order formalism equivalent of dimensionally reducing the Ricci scalar directly. And indeed, the $SL(2,\mathbb{R})$ particle-on-a-group action is equivalent to the Schwarzian action \cite{MTV}. Operator insertions on the other hand are not so simple.
\item 3d bulk gravity coupled to 3d CS theory leads to decoupled equations of motion because $T_{\mu\nu}^{CS}\equiv 0$. The only influence of the CS theory on the gravity part is in the definition of the total Hamiltonian: $H = H_{\text{grav}} + H_{CS}$ with contribution \eqref{csene}, which provides just a shift in the energy. This will indeed be observed below in section \ref{sect:u1}.
\end{itemize}

\subsection{Supersymmetric JT gravity theories}
\label{sect:susy}
The identification of the non-interacting gauge sector as a 2d BF theory can also be understood from supersymmetry as will be illustrated here.
Pure 3d gravity can be written as a $\mathfrak{sl}(2) \oplus \mathfrak{sl}(2)$ Chern-Simons theory. Similarly, Achucarro, Townsend and Witten demonstrated a long time ago that ($p$, $q$) 3d supergravity can be written as a $\mathfrak{osp}(p|2) \oplus \mathfrak{osp}(q|2)$ Chern-Simons theory \cite{Achucarro:1987vz}\cite{Witten:1988hc}. Dimensionally reducing these (super)gravity theories for the case $p=q$ leads to a $\mathfrak{osp}(p|2)$ 2d BF theory. \\
And indeed, as known since a long time \cite{Jackiw:1992bw}, JT gravity itself can be written as an $\mathfrak{sl}(2)$ BF theory:
\begin{equation}
\label{JTBF}
S_{JT} = \int \text{Tr}(\eta F),
\end{equation}
with $A = e_a P^a + \omega J$, field strength $F= dA + A \wedge A$ and $\eta = \eta_a P^a + \eta_3 J$ in terms of zweibein $e^a$ ($a=1,2$) and spin connection $\omega$. 

Supersymmetric generalization is now straightforward, as one just generalizes the gauge group from $\mathfrak{sl}(2)$ to either $\mathfrak{osp}(1|2)$ ($\mathcal{N}=1$) or $\mathfrak{osp}(2|2)$ ($\mathcal{N}=2$). In particular the $\mathcal{N}=2$ JT supergravity action may be written as \cite{Livine:2007dx}\cite{Astorino:2002bj}:
\begin{equation}
S_{JT}^{\mathcal{N}=2} = \int \text{STr} (\mathcal{E} F),
\end{equation}
in terms of the field strength $F = d\mathcal{A} + \mathcal{A} \wedge \mathcal{A}$, with the dilaton superfield $\mathcal{E}$ and superconnection $\mathcal{A}$, expanded into the $\mathfrak{osp}(2|2)$ generators as:
\begin{align}
\mathcal{E} &= \eta_a P^a + \eta_3 J + \phi^\alpha Q_\alpha + \tilde{\phi}^\alpha \tilde{Q}_\alpha + \chi B, \quad a=1,2, \quad \alpha = 1,2,\\
\mathcal{A} &= e_a P^a + \omega J + \psi^\alpha Q_\alpha + \tilde{\psi}^\alpha \tilde{Q}_\alpha + \xi B,
\end{align}
for three $\mathfrak{sl}(2)$ generators $P_a, J$, four fermionic generators $Q_\pm, \tilde{Q}_\pm$ and one additional $\mathfrak{u}(1)$ generator $B$. These eight generators satisfy an $\mathfrak{osp}(2|2)$ algebra whose explicit form can be found in the literature.\footnote{The $\mathfrak{osp}(1|2)$ BF theory would have just the 3 $\mathfrak{sl}(2)$ bosonic generators and 2 fermionic generators. Generally, the $\mathfrak{osp}(p|2)$ 2d BF theory has the bosonic algebra $\mathfrak{so}(p)\oplus \mathfrak{sl}(2)$.} \\
For simplicity, we set the cosmological constant zero here, as this does not influence the structure of the theory. In components, the action is
\begin{equation}
S_{JT}^{\mathcal{N}=2} = \int \left[\eta_a De^a + \eta_3 R + \eta \psi \wedge \tilde{\psi} + \chi (F + \psi \wedge \tilde{\psi}) + \tilde{\phi} D\psi + \phi D \tilde{\psi}\right].
\end{equation}
The piece coming from just the bosons is then
\begin{equation}
S_{JT}^{\mathcal{N}=2} \ni \int \eta_a De^a + \eta_3 R +\chi F,
\end{equation}
which is indeed bosonic JT gravity \eqref{JTBF} supplemented with a $\mathfrak{u}(1)$ BF theory $\int \chi F$. \\
Studying the $\mathcal{N}=2$ theory on its own would be interesting as this couples the gravitational and gauge sectors in the bulk. This is left for future work.

\section{Correlation functions in group models}
\label{sect:group}
We focus now on the boundary theories of the 3d Chern-Simons and 2d BF models. We will provide a prescription for computing correlation functions of the 1d particle-on-a-group theory, following the logic used in the Schwarzian theory in \cite{MTV} and in section \ref{sect:path}. We start by providing a general formalism starting from 2d Wess-Zumino-Witten (WZW) rational CFT and performing a double-scaling limit. Our main interest is again in computing the cylinder amplitude between vacuum branes. After that, we consider $U(1)$ and $SU(2)$ as two examples that will allow us to write down the generic correlation function using diagrammatic rules.

\subsection{General formalism}
\subsubsection{From 2d WZW to 1d particle-on-a-group}
Consider the 2d WZW system with path integral
\begin{equation}
\left\langle F(g(z,\bar{z}))\right\rangle = \frac{1}{Z}\int \left[\mathcal{D}g\right] F(g(z,\bar{z})) e^{-\frac{k}{16\pi}\int d^2 z \text{Tr}(g^{-1}\partial g g^{-1}\bar{\partial} g) + k \Gamma},
\end{equation}
for $g\in G$, integer level $k$, and with $\Gamma$ the Wess-Zumino term which will not be needed. An operator $F(g)$ is inserted, with $F$ a scalar-valued function on the group. As well-known, this theory enjoys invariance under a local group transformation $g \to g_1(z) g g_2(\bar{z})$. \\

Just as in Liouville theory, we focus on the moduli space of classical solutions of this theory to deduce the link between the 2d and 1d operators. This system has the classical solution $g(z,\bar{z}) = f(z)\bar{f}(\bar{z})$, with $f$ and $\bar{f}$ local group elements as well. \\
Inserting a brane at $z=\bar{z}$ (or $u=v$ in Lorentzian signature) imposes reflecting boundary conditions: 
\begin{equation}
\label{bdycond}
J(z) = \bar{J}(\bar{z}) \quad \Rightarrow \quad -\partial g g^{-1} = g^{-1}\bar{\partial}g,
\end{equation}
which, when translated into a condition on $f$, requires $\bar{f} = f^{-1}$. This boundary condition projects the symmetry onto its diagonal subgroup; the condition \eqref{bdycond} is preserved under the group transformation provided $g_1 = g_2^{-1}$. In terms of $f$, the symmetry transformation is now $f \to g_1 f$. \\
At the second boundary brane at $\sigma=\pi$, where $u=\tau+\pi, \, v=\tau-\pi$, one has $g = f(\tau+\pi)f^{-1}(\tau-\pi)$ which satisfies the boundary condition if $f$ is $2\pi$-periodic: $f(x + 2\pi) = f(x)$. Hence, after implementing the boundary conditions, the system is characterized by a single $2\pi$-periodic function $f$. \\
Just as with the Schwarzian theory, we imagine performing a change of field variables from $g$ to $f$. The transformation $g(z,\bar{z})=f(z)f^{-1}(\bar{z})$ has, in analogy with \eqref{GNtrans}, a redundancy in description: $f\, \sim \, f \gamma$ for $\gamma \in G$ any \emph{global} group element.
One can then identify a local WZW operator $F(g(z,\bar{z}))$ with a bilocal 1d operator as $z \to t_1$ and $\bar{z} \to t_2$. \\
Dimensionally reducing as in the Liouville/Schwarzian case, the WZW action itself immediately reduces to the particle-on-a-group action, the Wess-Zumino term $\Gamma$ vanishes upon dimensional reduction. \\

Hence the rational generalization of the Schwarzian story requires us to compute the 1d path integral over the group:
\begin{equation}
\label{1dint}
\frac{1}{Z}\int\displaylimits_{\tiny \left.\begin{array}{c}
G_{\text{local}}/G_{\text{global}} \\
f(t+2\pi)=f(t)
 \end{array}\right.} \hspace{-0.5cm}\left[\mathcal{D}f\right] F\left(f(t_1) f^{-1}(t_2)\right) e^{-\frac{kT}{16\pi} \int_{-\pi}^{\pi} dt \text{Tr}(f^{-1}\partial_t f)^2}.
\end{equation}
The periodicity of $2\pi$ can be changed into $\beta$ by rescaling the time coordinate as $t \to \frac{2\pi}{\beta}t$, which can alternatively be achieved by placing the branes at $\beta/2$ apart. Both the action and the operator insertions are left invariant under the \emph{global} group $f \to f \gamma$, but are not invariant under local transformations. This immediately generalizes the Schwarzian coset $\text{diff}S^1/SL(2,\mathbb{R})$ to the generic rational case as the right coset $G_{\text{local}}/G_{\text{global}}$. Taking into account the periodicity of $f$, this integration space is also written as the right coset of the loop group: $LG/G$, which is known to be a symplectic manifold. The resulting partition function could then be computed using the Duistermaat-Heckman (DH) theorem just as in the Schwarzian case \cite{Picken:1988ev}. Note that the transformation $f \to g_1 f$, $g_1 \in G$, is a symmetry of the action: it is the remnant of the WZW symmetry in 1d as remarked above. But it is not necessarily a symmetry of operator insertions and it isn't a gauge redundancy. \\
We did not work out the measure $\left[\mathcal{D}f\right]$ explicitly as in section \ref{sect:path}, but by general arguments this has to be the standard $\sqrt{G}$ measure of the group metric: $ds^2 = G_{\mu\nu}dx^\mu dx^\nu = \text{Tr} \left[g^{-1}dg \otimes g^{-1}dg\right]$. \\
The double scaling limit we take is $T\to 0$ and $k\to \infty$ with the product $kT \sim C$ held fixed proportional to a coupling constant $C$. We will be more specific about this below in section \ref{sect:sutwo}.\footnote{As for the Schwarzian case, reintroducing $\beta$ makes the constant $C$ have dimensions of length. The quantization of the level $k$ is immaterial in the double scaling limit.} The coupling constant $C$ allows us to explore the semi-classical regime of \eqref{1dint} at $C\to+\infty$. \\

Structurally the particle-on-a-group action is very similar to the Schwarzian action. The Lagrangian $L$ and Hamiltonian $H$ can be written as a particle moving on the group manifold as
\begin{equation}
L = \frac{C}{2} G_{\mu\nu}\dot{x}^\mu\dot{x}^\nu, \quad H = \frac{1}{2C}G^{\mu\nu}p_\mu p_\nu,
\end{equation}
making it clear that this action has $H=L$. The classical equations of motion are $\partial_t (f^{-1}\dot{f}) = 0$, identifying conserved currents $J(t) = J_a(t) \tau^a = f^{-1}\dot{f}$ with Casimir equal to the Hamiltonian (up to an irrelevant prefactor):
\begin{equation}
\text{Cas} \equiv \text{Tr}(J(t)J(t)) = J_a J_b \text{Tr}(\tau^a \tau^b) = \text{Tr}(f^{-1}\partial_t f)^2 \sim H = L.
\end{equation}

The quantization of a particle on a group manifold is in principle well-known (see e.g. \cite{Marinov:1979gm}). Consider for instance the partition function (without operator insertions), and ignore first the modding $f\sim f \gamma$ we wrote in \eqref{1dint}. Then this is manifestly the path integral rewriting of the Lorentzian partition function $\text{Tr} e^{-\beta H}$.  As mentioned above, the theory is invariant under $G\times G$ as $f(t) \to g_1 f(t) g_2$. Using operator methods, this can be used to prove that each energy-eigenvalue, with irrep label $j$, has a degeneracy of $(\text{dim j})^2$. As an example, the $SU(2)$ group manifold is just the three-sphere $S^3$, which has $SO(4) \simeq SU(2) \times SU(2)$ isometry, meaning an organization of the energy spectrum in $(2j+1)^2$ degenerate states. This can indeed also be seen explicitly for $SU(2)$ in \cite{Chu:1994hm}, and in the general case in \cite{Marinov:1979gm,Picken:1988ev}, both with operator methods and path integral methods. Thus
\begin{equation}
Z = \sum_j (\text{dim j})^2 \, e^{-\beta C_j}.
\end{equation}
Reintroducing the gauge-invariance $f\sim f \gamma$ in \eqref{1dint} merely requires gauge fixing the thermal path integral, which yields an overall factor of the (finite) group volume $(\text{vol } G)^{-1}$, which is included in the zero-temperatore entropy $S_0$ and dismissed. As mentioned above, this does however allow one to prove one-loop exactness of the path integral through the DH formula. The above expression is indeed what we will obtain in section \ref{sect:sutwo} below for $SU(2)$, and is readily generalized beyond that. We provide some more explicit formulas in appendix \ref{app:partgroup}.
 
\subsubsection{Cylinder amplitude}
Just as to get to the Schwarzian from Liouville in section \ref{sect:path}, we place two vacuum branes and consider the WZW amplitude on a cylinder between these vacuum branes (as earlier in Figure \ref{zzbrane}):
\begin{equation}
\label{amplii}
\left\langle \text{brane}_0\right| e^{- \tilde{T} H_{cl}} \left|\text{brane}_0\right\rangle,
\end{equation}
with $\tilde{T} =  2\pi^2/T$, the length of the cylinder in the closed channel when the circumference is fixed to $2\pi$. As well-understood, a boundary state $\left|a\right\rangle$ can be expanded into Ishibashi states as
\begin{equation}
\left|a\right\rangle = \sum_{i} \frac{S_a^i}{\sqrt{S_0^i}}|\hat{i}\rangle \hspace{-0.2em}\rangle.
\end{equation}
The sum ranges over all integrable representations of the Kac-Moody algebra $\hat{\mathfrak{g}}$, which in the $k\to+\infty$ limit becomes just all irreducible representations of the Lie algebra $\mathfrak{g}$. 
In the limit of interest where the length of the cylinder becomes much longer than its circumference, the Ishibashi states are themselves dominated by their zero-mode ($n=0$) states\footnote{All states obtained by acting with $J^{a}_{-n}$ on a primary state have non-trivial dependence on $\tau$, and are subdominant in the $T\to 0$ limit.}
\begin{equation}
\label{ishistate}
|\hat{i}\rangle \hspace{-0.2em} \rangle = \sum_{\mathbf{m}_i,n} \left|i,\mathbf{m}_i,n\right\rangle \otimes \left|i,\mathbf{m}_i,n\right\rangle \,\, \to \,\, \sum_{\mathbf{m}_i} \left|i,\mathbf{m}_i,n=0\right\rangle \otimes \left|i,\mathbf{m}_i,n=0\right\rangle.
\end{equation}
The Kac-Moody algebra reduces to the zero-mode Lie algebra. One can thus write for \eqref{amplii}:
\begin{equation}
\sum_{i,j}\sqrt{S_{0i}^*S_{0j}}\sum_{\mathbf{m}_i,\mathbf{m}_j}\left\langle i,\mathbf{m}_i\right| \delta_{ij} e^{- \beta C_j} \left|j,\mathbf{m}_j\right\rangle,
\end{equation}
in terms of the modular $S$-matrix and the Casimirs $C_i$ of the irreps. Including operator insertions in the middle, requires splitting the evolution into separate pieces and inserting complete sets of primaries around each such insertion. For instance, the two-point function of this system can be written as:
\begin{equation}
\sum_{i,j}\sqrt{S_{0i}^*S_{0j}}e^{-C_i \tau}e^{-C_j (\beta-\tau)} \sum_{\mathbf{m}_i,\mathbf{m}_j}\left\langle i,\mathbf{m}_i \right|F(g)\left| j,\mathbf{m}_j\right\rangle.
\end{equation}
The matrix element can e.g. be computed in configuration space as
\begin{equation}
\label{melement}
\left\langle i,\mathbf{m}_i \right|F(g) \left| j,\mathbf{m}_j\right\rangle = \int dg \left\langle i,\mathbf{m}_i\right|\left.g\right\rangle F(g) \left\langle g\right|\left.j,\mathbf{m}_j\right\rangle,
\end{equation}
which is the method we utilized for the Schwarzian theory in \cite{MTV}. \\

In the next two subsections we will consider the two simplest examples. The generalization to arbitrary compact groups will be obvious at the end. We will end up with a diagrammatic decomposition of the general correlator, analogously as in the Schwarzian case \cite{MTV}. Just as in that case, we remark that the resulting expression is non-perturbative in the coupling constant $C$: the diagrams just represent convenient packaging of the building blocks of the general expressions.

\subsection{Example: $U(1)$}
\label{sect:u1}
As a first example, let's take $U(1)$. We start with a direct evaluation of its correlators following the preceding discussion. Afterwards we will embed the theory into $\mathcal{N}=2$ Liouville and find the same answer. The latter serves as a further consistency check on the Schwarzian limit from supersymmetric versions of Liouville theory.

\subsubsection{Direct evaluation}
\begin{figure}[h]
\centering
\includegraphics[width=0.4\textwidth]{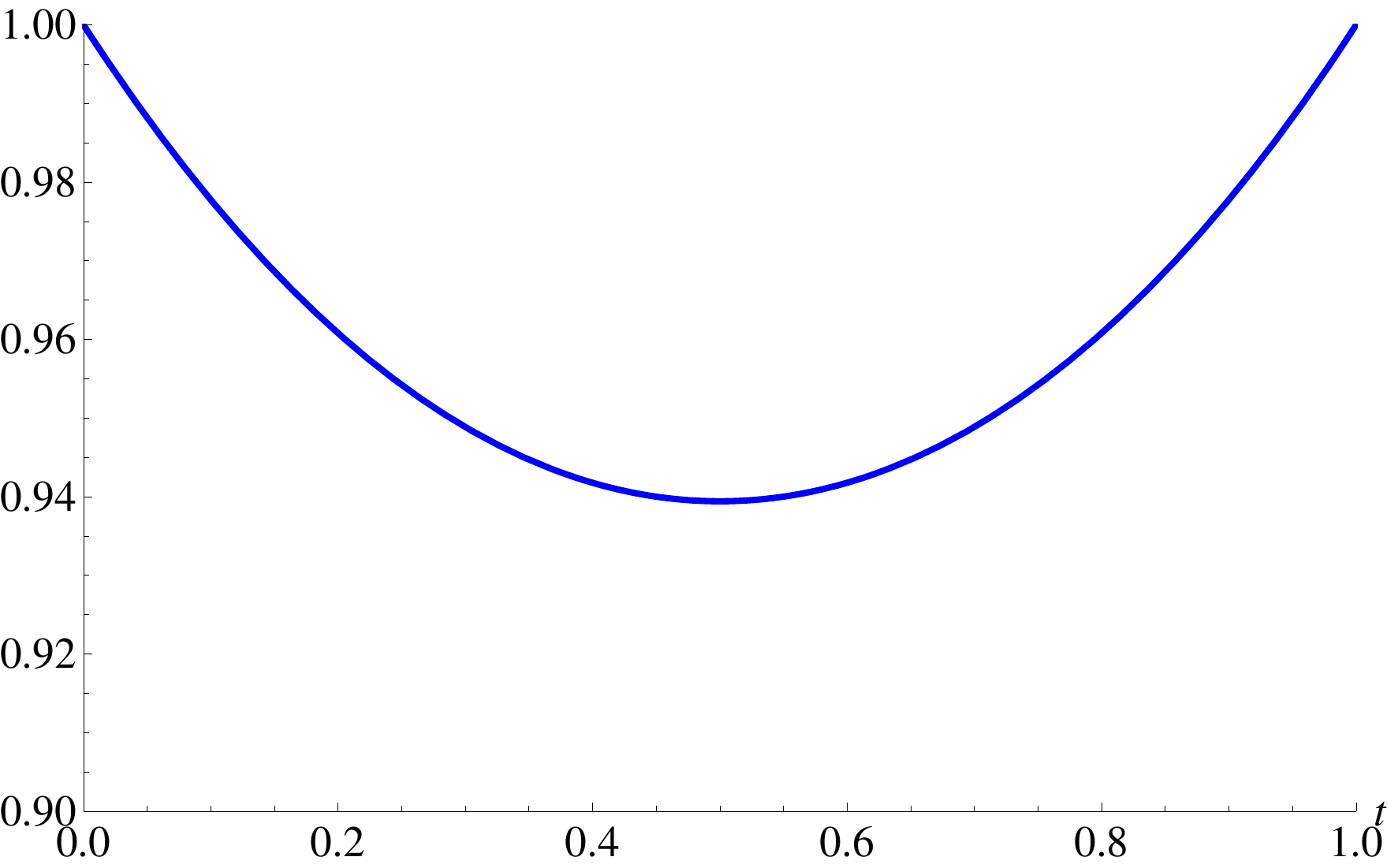}
\caption{Two-point function of $U(1)$ theory in units where $\beta=1$.}
\label{plotu1}
\end{figure}

Consider a free boson field $\phi$ in 2d with action $S = \int dudv \partial_u \phi \partial_v \phi$. The classical solution is given by
\begin{equation}
\phi(u,v) = \sigma(u) + \bar{\sigma}(v).
\end{equation}
Perfect reflection at $u=v$ and $u-v = 2\pi$ requires $\sigma = -\bar{\sigma}$ and $\sigma(u+2\pi) = \sigma(u)$. \\
Natural vertex operators are the exponentials:
\begin{equation}
V_Q = e^{iQ \phi(u,v)} = e^{iQ \sigma(u)} e^{-iQ \sigma(v)}.
\end{equation}
The classical moduli space is parametrized by a real periodic function $\sigma$, so the Schwarzian 1d limit entails:
\begin{equation}
\int \left[\mathcal{D}\phi\right] V_Q \hdots \, e^{-S} \to \int \left[\mathcal{D}\sigma\right] e^{iQ \sigma(t_1)} e^{-iQ \sigma(t_2)} \hdots \, e^{-\frac{1}{2}\int dt (\partial_t \sigma)^2}.
\end{equation}
In this particular case, the bilocal operator is just a product of two local operators. \\
Of course the resulting theory is free and immediately solvable. Consider e.g. a two-point correlator:
\begin{equation}
\left\langle e^{iQ \sigma(t_1)} e^{-iQ\sigma(t_2)}\right\rangle.
\end{equation}
The classical equation of motion for $\sigma$, including the operator insertions, is solved analogously as in the semi-classical regime of Liouville theory (and written here in Lorentzian signature):
\begin{equation}
\ddot{\sigma} = Q\delta(t-t_1) - Q\delta(t-t_2),
\end{equation}
hence $\dot{\sigma}$ increases by $Q$ at $t_1$ and decreases again to its original value at $t_2$. Thus the operators inject and extract charge, and $\dot{\sigma}$ represents the total charge in the system, as found earlier from the bulk perspective in section \ref{sect:BF}. The Gaussian path integral is readily computed as:
\begin{equation}
\label{u1int}
\frac{1}{Z}\int \left[\mathcal{D}\sigma\right]e^{i Q (\sigma_1-\sigma_2)}e^{-\int dt \dot{\sigma}^2} = \sqrt{\frac{\beta}{4\pi}}\int dq e^{-\frac{q^2}{4} \tau}e^{-\frac{(q-Q)^2}{4} (\beta-\tau)}.
\end{equation}
If the integral on the r.h.s. is truly an integral ranging from $-\infty$ to $+\infty$, one obtains:
\begin{equation}
e^{-\frac{Q^2\tau(\beta-\tau)}{4\beta}},
\end{equation}
which at $\beta\to+\infty$ asymptotes to $\to e^{-\frac{Q^2\tau}{4}}$. This, as we show below in \eqref{zerotnonab}, is the general result for any non-abelian group as well, with Casimir $Q^2/4$. This two-point function is of the shape as in Figure \ref{plotu1}.

\subsubsection{Interpretation in terms of $\mathcal{N}=2$ super-Schwarzian}
The $U(1)$-sector is relevant for e.g. the $\mathcal{N}=2$ super-Schwarzian. This is because it contains, in addition to the fermionic superpartners, also an additional bosonic field $\sigma$ that is identified with the above $U(1)$-sector. Here we demonstrate this directly. In the next paragraphs we will identify it from its $\mathcal{N}=2$ Liouville ancestor. \\

The bosonic piece of the super-Schwarzian action is the Schwarzian plus a free boson field $\sigma$ \cite{Fu:2016vas}:
\begin{equation}
\label{Sschaction}
S = C \int dt \left(-\left\{f,t\right\} + 2\dot{\sigma}^2\right).
\end{equation}
The relative coefficient was fixed by $\mathcal{N}=2$ supersymmetry. An $\mathcal{N}=2$ super-reparametrization of the invariant super-distance is given by the following expression:
\begin{equation}
\label{n2repara}
\frac{1}{\tau_1-\tau_2-\theta_1\bar{\theta}_2-\theta_2\bar{\theta}_1} \quad \to \quad 
\frac{\mathcal{D}_{\bar{\theta}_1} \bar{\theta}_1' \mathcal{D}_{\theta_2} \theta_2'}{\tau_1'-\tau_2'-\theta_1'\bar{\theta}_2'-\theta_2'\bar{\theta}_1'}.
\end{equation}
For a purely bosonic reparametrization,
\begin{equation}
\tau' = f(\tau), \quad \theta' = \rho(\tau) \theta, \quad \bar{\theta}' = \bar{\rho}(\tau) \bar{\theta}, \quad \text{with} \quad \rho\bar{\rho} = \dot{f}, \quad \rho/\bar{\rho} = e^{2i\sigma},
\end{equation}
the bosonic piece of (\ref{n2repara}) is given by
\begin{equation}
\frac{e^{i(-\sigma_1+\sigma_2)}\sqrt{\dot{f}_1\dot{f}_2}}{(f_1-f_2)}.
\end{equation}
This can be viewed as a simultaneous reparametrization $f(\tau)$ and gauge transformation $g(\tau) \equiv e^{i\sigma(\tau)}$ on the charged 1d operator $\mathcal{O} \to e^{i\sigma} \mathcal{O}$, as given in \eqref{susygreen}.

\subsubsection{Charged Schwarzian from $\mathcal{N}=2$ Liouville}
It is possible to obtain this theory directly from $\mathcal{N}=2$ Liouville theory. The $\mathcal{N}=2$ supersymmetric generalization of Liouville theory consists of the Liouville field $\phi$, the superpartners $\psi^\pm$ and $\bar{\psi}^{\pm}$ and a compact boson $Y$, forming the full supersymmetric multiplet. The central charge is $c=3+3\mathcal{Q}^2 = 3 + 3/b^2$. Details can be found in the literature, but will not be needed here.\footnote{Two convention schemes exist: we follow that of \cite{Eguchi:2003ik}. To go from the conventions of \cite{Ahn:2003tt} to those of \cite{Eguchi:2003ik}, one needs to set $b^2\to 2b^2$ and $2P^2\to P^2$.} \\

\begin{figure}[h]
\centering
\includegraphics[width=0.6\textwidth]{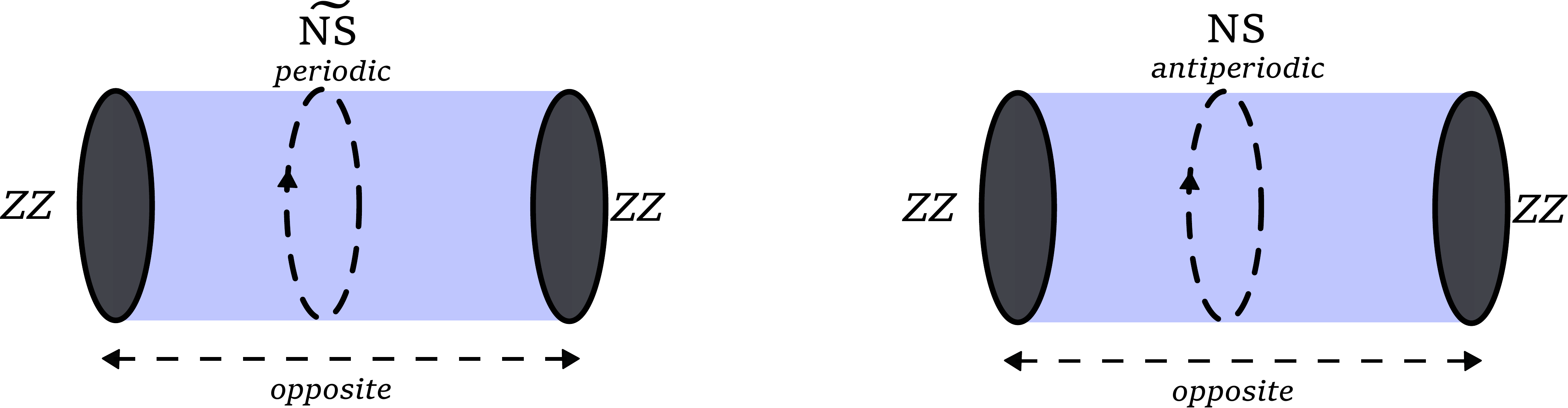}
\caption{Left: Cylinder with $\widetilde{NS}$ boundary conditions around the small circle, leading to supersymmetric quantum mechanics in 1d. Right: Cylinder with $NS$ boundary conditions around the small circle, leading to a removal of all fermions upon dimensional reduction.}
\label{fermiBCN2}
\end{figure}

Take this theory on the cylinder bounded by two ZZ-branes and consider imposing antiperiodic boundary conditions in $\mathcal{N}=2$ Liouville along the small circle (NS-sector) (Figure \ref{fermiBCN2}).

This leads to the removal of all fermionic degrees of freedom in the 1d theory, and retains only the Liouville field itself (leading to the Schwarzian) and the compact boson $Y$ (leading to the $U(1)$ theory). 
The analysis of section \ref{sect:path} can be repeated when adding the free boson $Y$. This leads to the additional 1d action in the Schwarzian limit:
\begin{equation}
S = \frac{C}{2} \int dt \, \dot{Y}^2,
\end{equation}
leading to the identification $Y=2\sigma$ to match with the super-Schwarzian field $\sigma$ in \eqref{Sschaction}.
The required building blocks of our story are readily available in the literature. $\mathcal{N}=2$ Liouville primary vertex operators in the $NS$ sector are of the form:
\begin{equation}
\label{vop}
V_{\ell,Q} = e^{\ell \phi}e^{i \frac{Q}{2} Y}, \quad \Delta = \frac{\ell}{2} - \frac{b^2}{2}\left(\ell^2-Q^2\right) \to \frac{\ell}{2},
\end{equation}
whereas Liouville states $\left|P,Q\right\rangle$ with charge $Q$ and Liouville momentum $P$ have weight:
\begin{equation}
\Delta = \frac{1}{8b^2} + \frac{P^2}{2} + \frac{b^2 Q^2}{2}.
\end{equation}
The $NS$ character for a primary with Liouville momentum $P=2bk$ and $U(1)$ charge $Q$ is given by:
\begin{equation}
\text{ch}^{NS}_{P,Q}(\tau,z) = q^{\frac{P^2}{2} + \frac{b^2Q^2}{2}} y^Q \frac{\theta_{00}(\tau,z)}{\eta^3} \,\, \to \,\, e^{-\beta(k^2+Q^2/4)}y^Q,
\end{equation}
in the large $\tau_2$-limit. The ZZ-brane wavefunction is determined by the modular $S$-matrix as:
\begin{equation}
\left|\Psi_{\text{ZZ}}(P,Q)\right|^2 = S_0^{P,Q} = \frac{b}{2}\frac{\sinh(\pi \frac{P}{b})\sinh(2\pi b P)}{\big| \cosh \pi \left( bP  + i b^2 Q\right)\big|^2} \,\, \to \,\, 2b^3\pi k \sinh 2 \pi k.
\end{equation}
The total vacuum character then has the small $T$-behavior:
\begin{align}
\chi_0(\tau = iT) \,\, \to \,\, &\int dQ\, d k \, k \, \sinh 2 \pi k e^{-\beta(k^2+Q^2/4)} y^Q \nonumber \\
&= \int dQ \int_{Q/2}^{+\infty} dE \sinh 2 \pi \sqrt{E-Q^2/4} \, e^{-\beta E} y^Q,
\end{align}
hence the density of states is identified as
\begin{equation}
\rho(E,Q) = \sinh 2 \pi \sqrt{E-Q^2/4}.
\end{equation}
The lack of a $\sim 1/\sqrt{E}$ divergence as $E \to 0$ is an indication of the lack of supersymmetry \cite{Kanazawa:2017dpd}. \\

Inserting one vertex operator \eqref{vop} in the ZZ-cylinder amplitude, we get:
\begin{align}
&\left\langle \mathcal{O}_{\ell,Q}(\tau_1,\tau_2)\right\rangle = \left\langle ZZ \right| e^{-H \tau} e^{\ell \phi} e^{i \frac{Q}{2} Y} e^{-H (\beta-\tau)}\left|ZZ\right\rangle = 
\int_{q_1^2/4}^{+\infty} dE_1 \int_{q_2^2/4}^{+\infty} dE_2 e^{-E_1 \tau} e^{-E_2 (\beta-\tau)} \nonumber \\
&\times \left\langle ZZ \right|\left.k_1,q_1\right\rangle \int d\phi dY \left\langle k_1,q_1\right|\left.\phi, Y\right\rangle e^{\ell \phi } e^{i \frac{Q}{2} Y}\left\langle \phi, Y\right|\left.k_2,q_2\right\rangle \left\langle k_2,q_2\right|\left.ZZ\right\rangle.
\end{align}
Let's compute this explicitly. The ZZ-brane wavefunction is given by
\begin{equation}
\psi_{ZZ}(E,q) = \left\langle E,q\right|\left.ZZ\right\rangle = \frac{2\pi i b \sqrt{E-q^2/4}}{\Gamma(1+2 i \sqrt{E-q^2/4})}.
\end{equation}
The minisuperspace limit of bulk $\mathcal{N}=2$ Liouville theory leads to a removal of all fermions, and the result is the Schr\"odinger equation:
\begin{equation}
(-\partial_\phi^2 - \partial_Y^2 + e^{2\phi})\psi(\phi,Y) = E \psi(\phi,Y),
\end{equation}
with $E$ the energy, solved by
\begin{equation}
\psi_{E,q}(\phi,Y) = \left\langle \phi, Y\right|\left.E,q\right\rangle = \frac{2}{\Gamma(-i\sqrt{E-q^2/4})}K_{i\sqrt{E-q^2/4}}(e^\phi)e^{i \frac{q}{2} Y}.
\end{equation}

The basic integral we need to compute is
\begin{equation}
\int dY \int d\phi e^{\ell \phi} e^{i \frac{Q}{2} Y}\psi_{k_1,q_1}^*(\phi,Y)\psi_{k_2,q_2}(\phi,Y).
\end{equation}
The $Y$-integral just gives $\delta(Q-q_1+q_2)$ and the $\phi$-integral is the same as in bosonic Liouville \cite{MTV}. So we end up with
\begin{align}
\int dq \int_{q^2/4}^{+\infty} dE_1 \int_{(q-Q)^2/4}^{+\infty} dE_2 e^{-E_1 \tau} &e^{-E_2 (\beta-\tau)} \sinh(2\pi\sqrt{E_1-q^2/4}) \sinh(2\pi\sqrt{E_2-(q-Q)^2/4}) \nonumber \\
&\times \frac{\Gamma(\ell/2 \pm i \sqrt{E_1-q^2/4} \pm i \sqrt{E_2-(q-Q)^2/4})}{\Gamma(\ell)}.
\end{align}
Shifting the energy variables by the charge, then leads to:
\begin{align}
\left\langle \mathcal{O}_{\ell,Q}(\tau_1,\tau_2)\right\rangle = &\int dq \int_{0}^{+\infty} dE_1 \int_{0}^{+\infty} dE_2 e^{-(E_1 +q^2/4)\tau} e^{-(E_2 +(q-Q)^2/4)(\beta-\tau)} \nonumber \\
&\times \sinh(2\pi \sqrt{E_1}) \sinh(2\pi \sqrt{E_2}) \frac{\Gamma(\ell/2 \pm i E_1 \pm i E_2)}{\Gamma(\ell)},
\end{align}
where now the energy variables $E_1$ and $E_2$ are only the energies of the Schwarzian subsystem, not the total energy. Factorization is now manifest, and the $q$-integral agrees indeed with \eqref{u1int}.\footnote{The computation in this section is done for $C=1/2$ \cite{MTV}, which in \eqref{Sschaction} indeed yields the correct prefactor in the action to agree with \eqref{u1int}.} \\
One can write a diagrammatic decomposition of a general correlator, as done in \cite{MTV}. The two-point correlator for instance is given diagrammatically as:
\begin{equation}
{\cal A}_2(k_i,q,\ell,Q, \tau_i) =  \begin{tikzpicture}[scale=0.61, baseline={([yshift=0cm]current bounding box.center)}]
\draw[thick] (0,0) circle (1.5);
\draw[thick] (-1.5,0) -- (1.5,0);
\draw[fill,black] (-1.5,0) circle (0.1);
\draw (0,2) node {\small $k_1, q$};
\draw (-2,0) node {\small $\tau_2$};
\draw[fill,black] (1.5,0) circle (0.1);
\draw (2,0) node {\small $\tau_1$};
\draw (0,-2) node {\small $k_2, q-Q$};
\draw (-0,.4) node {\small $\ell, Q$};
\end{tikzpicture},
\end{equation}
where each line contains also a conserved charge, next to the Schwarzian $SL(2,\mathbb{R})$-labels.

\subsection{Example: $SU(2)$}
\label{sect:sutwo}
\subsubsection{Partition function}
The vacuum character for $SU(2)_k$ on a cylinder of circumference $T$ and length $\pi$, transforms under an $S$-transformation as:
\begin{equation}
\label{su2vac}
\chi_0\left(\frac{iT}{2\pi}\right) = \sum_{j=0}^{k/2}S_{0j} \chi_j\left(i\frac{2\pi}{T}\right), \quad\,\, S_{0j} = \sqrt{\frac{2}{k+2}}\sin\left(\frac{\pi (2j+1)}{k+2}\right), \quad j=0,\frac{1}{2},1 \hdots \frac{k}{2},
\end{equation}
which can be evaluated in the $T\to 0$ limit using
\begin{equation}
\chi_j\left(i\frac{2\pi}{T}\right) \to (2j+1)e^{-\frac{4\pi^2}{T} h_j} = (2j+1)e^{-\tilde{T} (2h_j)} = (2j+1)e^{-\frac{4\pi^2}{T(k+2)} j(j+1)},
\end{equation}
where $h_j = \frac{j(j+1)}{k+2}$. The second equality expresses the character in terms of the closed channel with length $\tilde{T}=2\pi^2/T$. Keeping fixed $T(k+2) = 4\pi^2/\beta$, this becomes $(2j+1)e^{-\beta C_j}$ with the Casimir $C_j=j(j+1)$. The analogue of the Schwarzian double scaling limit is here that the level $k\to+\infty$ as $T\to 0$. The vacuum character \eqref{su2vac} finally becomes:
\begin{equation}
\label{pfunction}
Z(\beta) \, = \, \lim\limits_{T \to 0} \chi_0\left(\frac{iT}{2\pi}\right) \, = \, \sum_{j} \frac{\sqrt{2}\pi}{(k+2)^{3/2}} (2j+1)^2 e^{-\beta C_j} \, = \, \sum_{j} S_{00} \, (2j+1)^2 e^{-\beta C_j},
\end{equation}
which, up to normalization constants, is a discrete quantum system with Hamiltonian = Casimir, and with the dimension of the irreps as density of states: $\rho(j,m) = \text{dim j} = 2j+1$. Note that the sum ranges over both integers and half-integers. \\
As in the Schwarzian theory, the prefactor can be written in terms of a ground state entropy as $e^{S_0}$, and requires regularization by taking finite $k$. In this case, the prefactor is just $S_{00}$ which goes to zero as $k\to\infty$. This prefactor will cancel in correlation functions and is hence irrelevant for our computations; we drop it from here on. \\
At low temperatures, only the vacuum contributes and $Z \to 1$. At high temperatures, the sum can be replaced by an integral and $Z \to \frac{2\sqrt{\pi}}{\beta^{3/2}}$. Alternatively, the expression \eqref{pfunction} is readily Poisson-resummed.

For a general Kac-Moody algebra $\hat{\mathfrak{g}}$, it is well-known that the $S_{0j}$ elements in the modular $S$-matrix carry information about the quantum dimension $d_j$ of the integrable representation $\hat{j}$, and this reduces to the dimension in the classical ($k\to\infty$) limit:
\begin{equation}
d_j = \frac{S_{0j}}{S_{00}} \, \to \, \text{dim } j.
\end{equation}
It is instructive to recompute $Z(\beta)$ from the closed channel:
\begin{align}
\left\langle \text{brane}_0\right| e^{- \tilde{T} H } \left|\text{brane}_0\right\rangle = \sum_{i,j}\sqrt{S_{0i}^*S_{0j}}\,\, \langle \hspace{-0.2em} \langle \hat{i}|e^{- \tilde{T} H}|\hat{j}\rangle\hspace{-0.2em}\rangle \, \to \, S_{00}\sum_{i,j}\text{dim i}\text{ dim j} \,\, \delta_{ij} e^{-\beta C_j},
\end{align}
using the Ishibashi states in the $k\to\infty$ limit \eqref{ishistate}:
\begin{equation}
\label{ishib}
| \hat{j} \rangle\hspace{-0.2em}\rangle \, \to \, \sum_{m =-j}^{j} \left|j,m\right\rangle.
\end{equation}

\subsubsection{Correlation Functions}
Next we proceed by computing correlators of the $SU(2)$ theory. Instead of evaluating configuration space integrals, we will compute the matrix element \eqref{melement} directly using group theory as follows. General operator insertions $F(g)$ are all built from the field $g(z)$, so we can organize them into tensor operators $\mathcal{O}_{J,M\bar{M}}$ transforming in an irreducible representation of $G$, essentially by using the Peter-Weyl theorem. In the double scaling limit, one finds the bi-local operators:
\begin{equation}
F(g) \, \to \, F\left(f(t_1)f^{-1}(t_2)\right) = \sum_{J,M,\bar{M}} c^{J,M,\bar{M}} \mathcal{O}_{J,M\bar{M}}.
\end{equation}
The resulting elementary bilocal operator $\mathcal{O}_{J,M\bar{M}}$ will turn out to be identifiable with the matrix element:
\begin{equation} 
\mathcal{O}_{J,M\bar{M}} \equiv \left[f(t_1)f^{-1}(t_2)\right]_{M\bar{M}} = \left[R_J(f(t_{1}))\right]_{M\alpha}\left[R_J(f^{-1}(t_2))\right]^{\alpha}_{\bar{M}},
\end{equation}
for the group element $f$ in the spin-$J$ representation $R_J$. For a general operator $\mathcal{O}_{J,M\bar{M}}$ transforming both in the holomorphic and antiholomorphic sector as a tensor operator, a doubled Wigner-Eckart theorem holds:
\begin{equation}
\label{melA}
\left\langle j_1m_1\bar{m}_1\right|\mathcal{O}_{J,M\bar{M}}\left|j_2m_2\bar{m}_2\right\rangle = C_{j_1m_1,j_2m_2;J-M}C_{j_1\bar{m}_1,j_2\bar{m}_2;J-\bar{M}} A_{j_1j_2J},
\end{equation}
in terms of two Clebsch-Gordan (CG) coefficients and a reduced matrix element $A_{j_1j_2J}$. Note that a reordering of the arguments of the CG coefficients has been performed, resulting in some $j$-dependent factors that are absorbed into the reduced matrix element, see appendix \ref{app:tech} for details. The appearance of two Clebsch-Gordan coefficients will be crucial in what follows. \\
To determine the reduced matrix element $A_{j_1j_2J}$, one can evaluate this expression for any choice of the $m$'s. \\

We will determine it below for $SU(2)$, and conjecture that for a general group $G$ for irreducible representations $\lambda_1$, $\lambda_2$ and $\lambda$, it equals
\begin{equation}
\label{Aconj}
\boxed{
A_{\lambda_1\lambda_2\lambda} = \sqrt{\frac{\text{ dim }\lambda_1 \text{ dim }\lambda_2}{\text{ dim }\lambda}}}.
\end{equation}

The $SU(2)_k$ OPE coefficient was written down in \cite{Gaberdiel:2007vu}, and is for the case $m_2=\bar{m}_2 = j_2$ and $m_3=\bar{m}_3 = J$
\begin{equation}
\label{stopeq}
\left\langle j_1m_1\bar{m}_1\right|\mathcal{O}_{J,M\bar{M}}\left|j_2m_2\bar{m}_2\right\rangle = \left\langle \Phi_{j_1 m_1 \bar{m}_1} \Phi_{J, M \bar{M}} \Phi_{j_2 m_2 \bar{m}_2}\right\rangle = D_{j_1J}^{j_2},
\end{equation}
for fusing $j_1$ and $j_2$ into $J$. In the large $k$ limit, this is given explicitly as
\begin{equation}
D_{j_1J}^{j_2} \, \to \, \sqrt{(2j_1+1)(2j_2+1)(2J+1)}\frac{\Gamma(2j_2+1)\Gamma(2J+1)}{\Gamma(j_1+j_2+J+2)\Gamma(j_2+J-j_1+1)}.
\end{equation}
On the other hand, the $SU(2)$ Clebsch-Gordan coefficient for combining $j_1$ and $j_2$ into $J$ equals
\begin{align}
C_{j_1m_1;j_2j_2;J-J} = \sqrt{\frac{(2J+1)\Gamma(2J+1)\Gamma(2j_2+1)}{\Gamma(j_1+j_2+J+2)\Gamma(J+j_2-j_1+1)}}(-)^{J-j_1-j_2}\delta_{\sum_i m_i}.
\end{align}
Some details of these computations are given in appendix \ref{app:tech}. We obtain the ratio
\begin{equation}
\frac{D_{j_1j_2}^{J}}{C_{j_1m_1;j_2j_2;J-J}^2} = \sqrt{\frac{(2j_1+1)(2j_2+1)}{2J+1}},
\end{equation}
identifying the reduced matrix element in \eqref{melA} as
\begin{equation}
A_{j_1j_2J} = \sqrt{\frac{(2j_1+1)(2j_2+1)}{2J+1}},
\end{equation}
which indeed suggests the general form \eqref{Aconj}. \\

The matrix element in the double scaling limit (and with the normalization \eqref{ishib}) is then written by the Wigner-Eckart theorem as
\begin{equation}
\label{melwe}
\left\langle \hspace{-0.2em}\left\langle j_1\right.\right| \mathcal{O}_{J,M\bar{M}}\left|\left.j_2\right\rangle\hspace{-0.2em}\right\rangle \, \to \, \sum_{m_1,m_2} C_{j_1m_1,j_2m_2;J-M}^2 \, A_{j_1j_2J} \, \delta_{M\bar{M}},
\end{equation}
in terms of the Clebsch-Gordon coefficients and the reduced matrix element. Only operators that are left-right symmetric can connect the two Ishibashi states, yielding the Kronecker-delta. The sum over CG coefficients squared is just the fusion coefficient:
\begin{equation}
\sum_{m_1,m_2} C_{j_1m_1,j_2m_2;J-M}^2 = N_{j_1j_2}^{J}.
\end{equation} 
It equals 1 by unitarity of the CG-matrix, and can connect only states satisfying the triangle inequality. The formula \eqref{melwe} is a classical limit of a formula recently derived by Cardy in \cite{Cardy:2017ufe} (derived there for diagonal minimal models) where the Ishibashi matrix element is written as
\begin{equation}
\left\langle \hspace{-0.2em} \left\langle j_1\right.\right| e^{-\tau H}\mathcal{O}_{J,MM}e^{-\tau H}\left|\left.j_2\right\rangle\hspace{-0.2em}\right\rangle = \left(\frac{\pi}{2\tau}\right)^{\Delta_J} (S_0^0)^{1/2} \sqrt{\frac{S_{0}^{j_1} S_0^{j_2}}{S_0^J}} N_{j_1j_2}^{J} ,
\end{equation}
for a (diagonal) primary operator $\mathcal{O}_{J,MM}$. The Euclidean propagators $e^{-\tau H}$ and the first factors on the r.h.s. can be viewed as regularization artifacts of the Ishibashi states to render them normalizable. We conjecture this formula and its classical limit hold for any rational CFT. In any case, we have illustrated it explicitly for $SU(2)_k$ which is the relevant symmetry group for e.g. $\mathcal{N}=4$ super-Schwarzian systems (see e.g. \cite{Aoyama:2018lfc}). \\

The normalization of the intermediate operator $\mathcal{O}_{J,M\bar{M}}$ has been fixed above by the 2d CFT state-operator correspondence in \eqref{stopeq}. There is however a more convenient normalization for the 1d theory, by taking the operator and the $SL(2)$-field $\Phi_{J,M\bar{M}}$ to be instead related as
\begin{equation}
\label{normalization}
\mathcal{O}_{J,M\bar{M}} = \frac{1}{\sqrt{\text{dim J}}}\Phi_{J,M\bar{M}},
\end{equation}
which we now adopt. \\

\begin{figure}[h]
\centering
\includegraphics[width=0.4\textwidth]{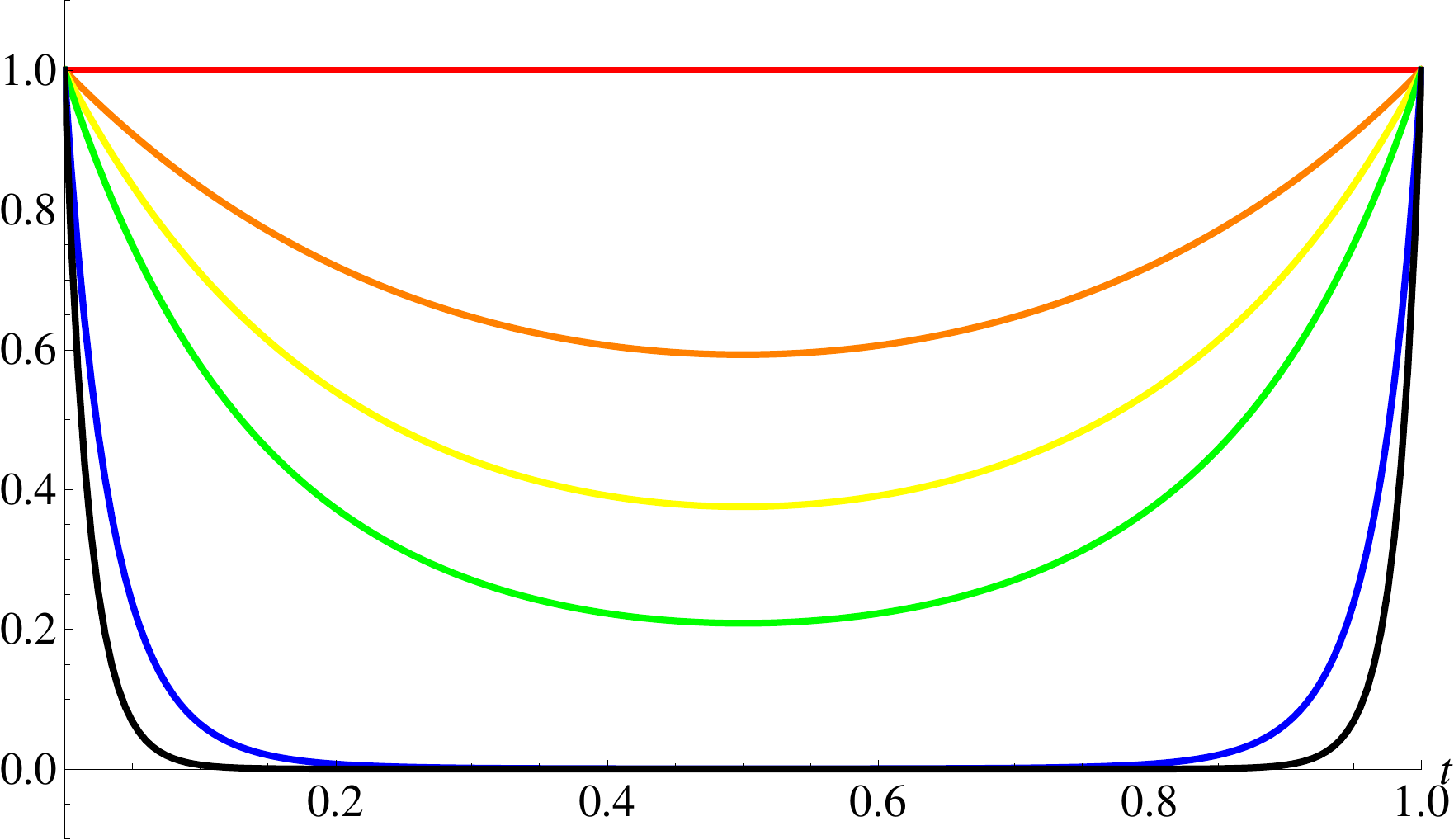}
\caption{Two-point function of $SU(2)$ theory \eqref{2ptsu2} in units where $\beta=1$ for several values of $J$: $J=0$ (red), $J=1$ (orange), $J=3/2$ (yellow), $J=2$ (green), $J=5$ (blue), $J=7$ (black).}
\label{su2draw}
\end{figure}

Higher-point functions can now be deduced analogously, and we arrive at a diagrammatic decomposition of a general correlation function, where one sums over all intermediate representation labels using
\begin{equation}
\sum_{j_i,m_i} \text{dim j}_i \,\, {\cal A}(j_i,m_i;\tau_i),
\end{equation}
and where the momentum amplitude ${\cal A}(j_i,m_i;\tau_i)$ is computed using the Feynman rules:
\bea 
\label{frules}
\begin{tikzpicture}[scale=0.57, baseline={([yshift=-0.1cm]current bounding box.center)}]
\draw[thick] (-0.2,0) arc (170:10:1.53);
\draw[fill,black] (-0.2,0.0375) circle (0.1);
\draw[fill,black] (2.8,0.0375) circle (0.1);
\draw (3.4, 0) node {\footnotesize $\tau_1$};
\draw (-0.7,0) node {\footnotesize $\tau_2$};
\draw (1.25, 1.6) node {\footnotesize $j m$};
\draw (6.5, 0) node {$\raisebox{6mm}{$\ \ =\ \ e^{-\spc \spc C_{j} \spc (\tau_2-\tau_1)}$}$};
\end{tikzpicture}, ~~~~~~~~~~\ \ \begin{tikzpicture}[scale=0.7, baseline={([yshift=-0.1cm]current bounding box.center)}]
\draw[thick] (-.2,.9) arc (25:-25:2.2);
\draw[fill,black] (0,0) circle (0.08);
 \draw[thick](-1.5,0) -- (0,0);
\draw (.5,-1) node {\footnotesize $\textcolor{black}{j_2m_2}$};
\draw (.5,1) node {\footnotesize$\textcolor{black}{j_1m_1}$};
\draw (-1,.3) node {\footnotesize$\textcolor{black}{JM}$};
\draw (3,0.1) node {$\mbox{$\ =\  \, \gamma_{j_1m_1,j_2m_2,JM}\spc .$}$}; \end{tikzpicture}\ 
\eea
The vertex is essentially the Clebsch-Gordan coefficient, but it can be written more symmetrically in terms of the $3j$-symbol:
\begin{equation}
\threej{j_1}{m_1}{j_2}{m_2}{j_3}{m_{3}} = \frac{(-)^{j_1-j_2-m_3}}{\sqrt{2j_3+1}}C_{j_1m_1,j_2m_2;j_3-m_3},
\end{equation}
as
\begin{equation}
\gamma_{j_1m_1,j_2m_2,JM}^2 = \threej{j_1}{m_1}{j_2}{m_2}{J}{M}^2.
\end{equation}
The CG coefficients determine the fusion of the representations at each double vertex. \\
Hence, we obtain finally for the 1d two-point function ($\mathcal{O}_{J,M}:=\mathcal{O}_{J,MM}$):
\begin{equation}
\left\langle \mathcal{O}_{J,M} \right\rangle = \sum_{j_1,j_2,m_1,m_2} \text{dim j}_1\text{ dim j}_2 \,\, {\cal A}_2(j_i,m_i,J, M,\tau_i),
\end{equation}
with the amplitude $\mathcal{A}_2$ diagrammatically:
\bea
{\cal A}_2(j_i,m_i,J, M,\tau_i)\  \is \  \ \ \begin{tikzpicture}[scale=0.61, baseline={([yshift=0cm]current bounding box.center)}]
\draw[thick] (0,0) circle (1.5);
\draw[thick] (-1.5,0) -- (1.5,0);
\draw[fill,black] (-1.5,0) circle (0.1);
\draw (0,2) node {\small $j_1m_1$};
\draw (-2,0) node {\small $\tau_2$};
\draw[fill,black] (1.5,0) circle (0.1);
\draw (2,0) node {\small $\tau_1$};
\draw (0,-2) node {\small $j_2m_2$};
\draw (-0,.4) node {\small $JM$};
\end{tikzpicture}
\eea
Combining everything we arrive at:
\begin{equation}
\boxed{
\left\langle \mathcal{O}_{J,M} \right\rangle = \frac{1}{Z(\beta)} \sum_{j_1,j_2,m_1,m_2} \text{dim j}_1 \text{ dim j}_2 \, e^{-C_{j_1} \tau}e^{-C_{j_2} (\beta-\tau)}  \threej{j_1}{m_1}{j_2}{m_2}{J}{M}^2},
\end{equation}
which, for the particular case of the two-point function, can be written fully in terms of the integer fusion coefficients $N_{j_1j_2}^{J}$:
\begin{equation}
\label{2ptsu2}
\left\langle \mathcal{O}_{J,M} \right\rangle \to \frac{1}{Z(\beta)} \sum_{j_1,j_2}\text{dim j}_1 \text{ dim j}_2 \,\, e^{-C_{j_1} \tau}e^{-C_{j_2} (\beta-\tau)} \frac{N_{j_1j_2}^{J}}{\text{dim J}}.
\end{equation}
This immediate simplification only occurs for the two-point function. \\
Just as for $U(1)$, this correlator is finite as $\tau \to 0$. The qualitative shape of the correlator is similar to the $U(1)$-case. Some examples are drawn in Figure \ref{su2draw}.
Our choice of normalization \eqref{normalization} ensures that $\left\langle \mathcal{O}_{J,M}(\tau=0) \right\rangle = 1$. As a check, some simplifying limits can be taken. At zero temperature, $C_{j_2}=0$, so $j_2=0$, and $J=j_1$. So
\begin{equation}
\label{zerotnonab}
\left\langle \mathcal{O}_{J,\mathbf{M}} \right\rangle_{\beta \to \infty} \to e^{-C_J \tau}.
\end{equation}
When $J=0$ (insertion of the identity operator), $j_1=j_2$ and one finds $\left\langle \mathcal{O}_{0,0} \right\rangle = 1$, confirming the overall normalization of \eqref{2ptsu2}. \\ 

The partition function $Z(\beta)$ itself \eqref{pfunction} is also directly computed using the Feynman diagram decomposition:
\bea
Z(\beta)\, = \sum_{j,m} \text{dim j} \,\, e^{-C_j \beta} = \sum_{j}(\text{dim j})^2 \, e^{-C_j \beta} =  \raisebox{1mm}{ \begin{tikzpicture}[scale=0.6, baseline={([yshift=0cm]current bounding box.center)}]
\draw[thick] (0,0) circle (1.5);
\end{tikzpicture}}
\eea
The time-ordered four-point function is drawn as
\beq
{\cal A}_4\bigl(j_i,m_i,J_i,M_i,\tau_i\bigr)\ =  \ \,\begin{tikzpicture}[scale=0.65, baseline={([yshift=0cm]current bounding box.center)}]
\draw[thick] (0,0) circle (1.5);
\draw[thick] (1.3,.7) arc (300:240:2.6);
\draw[thick] (-1.3,-.7) arc (120:60:2.6);
\draw[fill,black] (-1.3,-.68) circle (0.1);
\draw[fill,black] (1.3,-.68) circle (0.1);
\draw[fill,black] (-1.3,0.68) circle (0.1);
\draw[fill,black] (1.3,0.68) circle (0.1);
\draw (2.3,0) node {\footnotesize $j_3m_3$};
\draw (-2.3,0) node {\footnotesize  $j_3m_3$};
\draw (0,.8) node {\footnotesize $J_1M_1$};
\draw (0,-.8) node {\footnotesize $J_2M_2$};
\draw (0,1.88) node {\footnotesize  $j_1m_1$};
\draw (0,-1.88) node {\footnotesize  $j_2m_2$};
\draw (-1.75,-.75) node {\footnotesize $\tau_3$};
\draw (-1.75,.75) node {\footnotesize $\tau_2$};
\draw (1.75,-.75) node {\footnotesize $\tau_4$};
\draw (1.75,.75) node {\footnotesize $\tau_1$};
\end{tikzpicture}\eeq
and is given by the expression:
\begin{align}
\left\langle \mathcal{O}_{J_1,M_1} \mathcal{O}_{J_2,M_2}\right\rangle &= \sum_{j_i,m_i} \text{dim j}_1 \text{ dim j}_2 \text{ dim j}_3 \,\, {\cal A}_4(j_i,m_i,J_i, M_i,\tau_i) \nonumber \\
&=\frac{1}{Z(\beta)} \sum_{j_i,m_i} e^{-C_{j_1} (\tau_2-\tau_1)}e^{-C_{j_2} (\tau_4-\tau_3)}e^{-C_{j_3}(\beta-\tau_2+\tau_3-\tau_4+\tau_1)} \\
&\hspace{-1.0cm}\times\text{dim j}_1 \text{ dim j}_2\text{ dim j}_3 \threej{j_1}{m_1}{j_3}{m_3}{J_1}{M_1}^2\threej{j_3}{-m_3}{j_2}{m_2}{J_2}{M_2}^2. \nonumber
\end{align}
Note that as $\beta \to \infty$, this four-point function factorizes in two zero-temperature two-point function, coming from the clustering principle, and the dependence on only two independent time differences, just as happens in the Schwarzian case \cite{MTV}. \\

This construction is immediately generalized to arbitrary compact groups $G$, and leads to the rules as given in section \ref{sect:intro}. \\

The braiding and fusion matrices, which are given by $q$-deformed $6j$-symbols of the group $G$ \cite{PT}, become the classical $6j$-symbol of the group $G$. As emphasized for the Schwarzian case in \cite{MTV}, this quantity is used to swap the operator ordering and reach specific out-of-time ordered (OTO) correlators of interest, dual to shockwave interactions in the gravitational case \cite{Jackson:2014nla}. For the Schwarzian theory, we find the precise semi-classical (large $C$) shockwave expressions of \cite{Shenker:2014cwa,Maldacena:2016upp} starting from the exact OTO correlators in \cite{LMTV}. We leave a more detailed discussion to future work.

\section{Concluding remarks}
\label{sect:conc}
In this work, we presented more evidence and extensions to the link between 2d Liouville theory and the 1d Schwarzian theory. We believe this is the most natural way to look at the Schwarzian theory. The first half of this paper focussed on the Liouville path integral directly, where we emphasized the relevance of the parametrization of Gervais and Neveu in this context. \\
We further extended the AdS$_2$ argument for preferred coordinate frames of \cite{Almheiri:2014cka,Maldacena:2016upp,Engelsoy:2016xyb} to the case of gauge theories and preferred gauge transformations. \\
In the second half of this work, we demonstrated that the Schwarzian limit is only a special (irrational) case of the simpler case of rational compact models. All of these geometric theories have the property that the Hamiltonian, Lagrangian and Casimir coincide, and that local operators in 2d CFT become bilocal operators in 1d QM in a double-scaling limit. We produced correlation functions from the 2d WZW perspective, although our analysis was not entirely rigorous as we used the generalization of the prescription of \cite{MTV}. It would be an improvement to complement this with a path-integral analysis as in section \ref{sect:path} for the rational theories as well, including the measure in the path integral. This is left to future work. \\
Nonetheless, we deduced expressions for time-ordered correlators and provided diagrammatic rules. Out-of-time-ordered correlators can also be studied and require introducing $6j$ symbols to swap internal lines in diagrams. It would be particularly interesting to link this to results on OTO-correlators in rational 2d CFT, as in e.g. \cite{Caputa:2016tgt}. \\
These theories also seem to be related to group field theories, utilized in the spinfoam formulation of LQG, which in turn seem to be related to the tensor models of e.g. \cite{Witten:2016iux,Klebanov:2016xxf}. \\
A very interesting extension to study deeper would be to understand $\mathcal{N}=2$ Liouville theory in the $\widetilde{NS}$-sector, which allows one to connect to the 1d $\mathcal{N}=2$ super-Schwarzian theory. The latter contains non-trivial interactions between the $Y$-boson and the Liouville field $\phi$ itself. However, technical obstructions appear to be present when analyzing the mini-superspace regime and performing the $\phi$-integrals directly in coordinate space. We hope to come back to this problem in the future. \\
The structure present in the rational theories, suggests the Schwarzian three-point vertex $\gamma_\ell(k_1,k_2)$ should also be interpretable as a $3j$ symbol of $SL(2,\mathbb{R})$ with 1 discrete and 2 continuous representations. If this can be made more explicit, then the generalizations to the supersymmetric Schwarzian correlators can be conjectured to hold in terms of $3j$ and $6j$ symbols of OSp$(1|2)$ and OSp$(2|2)$ for $\mathcal{N}=1$ and $\mathcal{N}=2$ super-Schwarzian theories respectively, without resorting to the coordinate space evaluation of the Liouville integrals as mentioned above. \\
We will make the link between $SL(2,\mathbb{R})$ BF theory and the Schwarzian explicit in upcoming work, using a complementary bulk holographic perspective on bilocal correlation functions in terms of boundary-anchored Wilson lines in BF theory. This was already hinted at in \cite{Blommaert:2018oue}. \\
A further question is whether anything can be learned for 4d gauge theories, as 2d boundary Liouville/Toda CFT was demonstrated in an AGT context in \cite{LeFloch:2017lbt} to be linked to (a certain subclass of) these. Taking the double scaling limit should have an analogue in 4d gauge theories. \\
One of the holographic successes of the Schwarzian theory is a correct prediction of the Bekenstein-Hawking entropy of the JT black holes \cite{Maldacena:2016upp,Engelsoy:2016xyb}. Within the Liouville framework, it arises fully from the modular $S$-matrix as $S_{BH} = \log S_0^p$. On the other hand, it was found in \cite{McGough:2013gka} that the topological entanglement entropy in 2d irrational Virasoro CFT matches the Bekenstein-Hawking entropy for 3d BTZ black holes: $S_{BH} = \log S_0^{p+}S_{0}^{p-}$. It would be interesting to utilize the 2d/1d perspective to shed more light on some of the puzzles that appear in 3d gravity and its relation to 2d Liouville dynamics.

\begin{center}
{\bf Acknowledgements}
\end{center}

\vspace{-2mm}
I am deeply grateful to A. Blommaert, N. Callebaut, H. T. Lam, G. J. Turiaci and H. L. Verlinde for numerous discussions and questions that greatly benefitted this work. The author acknowledges financial support from the Research Foundation-Flanders (FWO Vlaanderen).


\begin{appendix}

\section{Virasoro coadjoint orbits and Liouville branes}
\label{app:coadjoint}

\begin{figure}[h]
\centering
\includegraphics[width=0.5\textwidth]{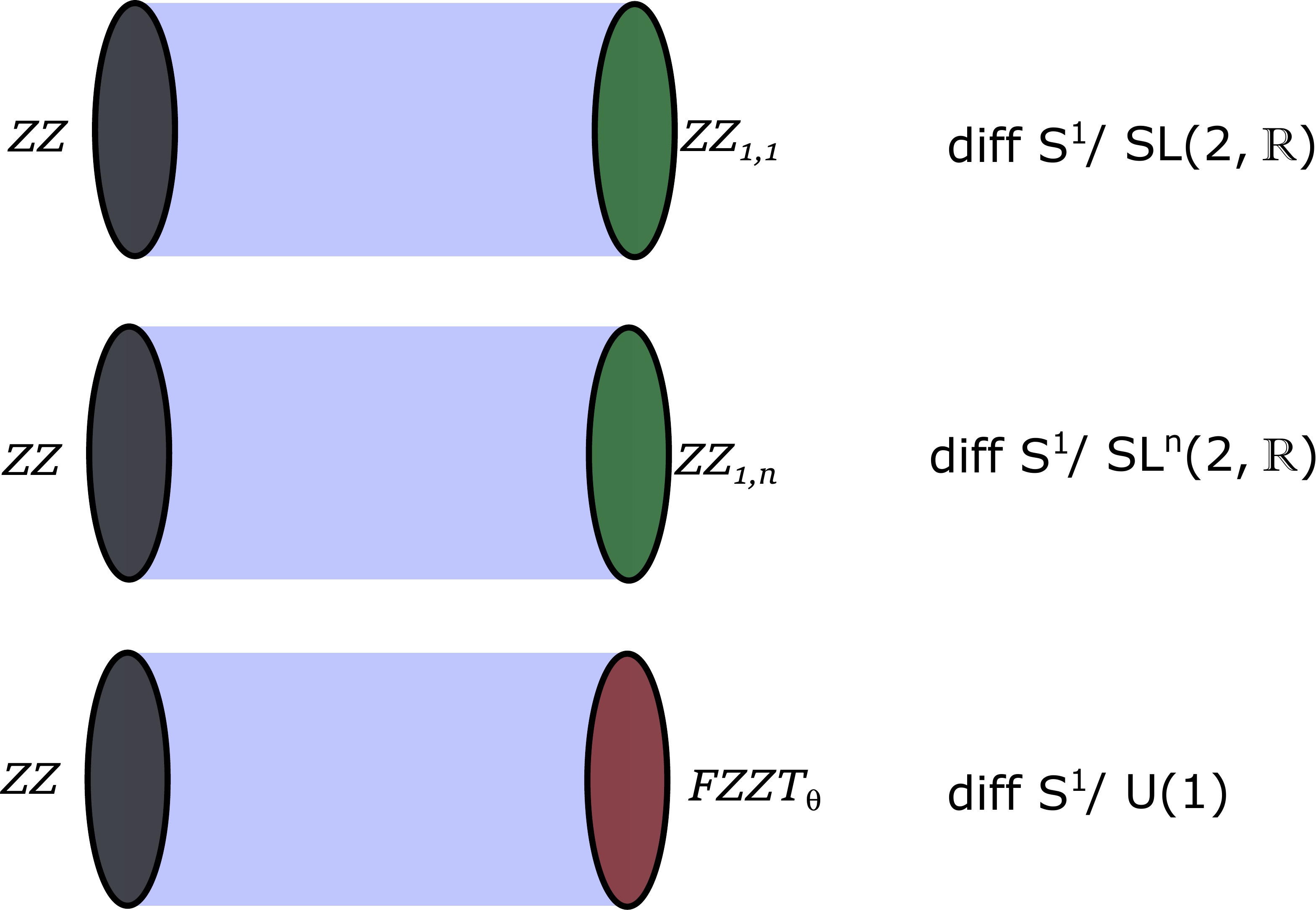}
\caption{Other brane configurations. Top: the ZZ-ZZ system. Middle: the ZZ-ZZ${}_{1,n}$ system. Bottom: the ZZ-FZZT system.}
\label{generalize}
\end{figure}

It is instructive to generalize the construction in section \ref{sect:path} to more general branes, and make the link with the Alekseev-Shatashvili geometric action and the coadjoint orbits of the Virasoro group more explicit. \\
W.l.g. we keep the left brane fixed as a ZZ-brane. We first present the 2d case, and discuss the Schwarzian limit at the very end only. We can now list the possible generalizations in Figure \ref{generalize}.\\

In section \ref{sect:path}, if one replaces the brane at $\sigma = \pi$ by a ZZ${}_{1,n}$ brane \cite{Zamolodchikov:2001ah}, one can use the same definition of $a$ and $b$, but now take the boundary condition as $\left.a=b+2\pi n\right|_{\sigma=\pi}$ for $n>1$. This gives singularities in the Liouville field $\phi$ also in between both branes. Setting $f \to n f$, the periodicity of $f$ is restored, and one obtains again a $\text{diff} S^1/SL(2,\mathbb{R})$ theory. 
Gervais and Neveu \cite{Gervais:1981gs,Gervais:1982nw,Gervais:1982yf,Gervais:1983am} considered boundary conditions that in modern parlance would be called FZZT branes \cite{Fateev:2000ik}. One can directly implement an FZZT brane at $\sigma=\pi$ by requiring instead $\left.a=b+2\pi\theta\,\right|_{\sigma=\pi}$, with $r^2 = \cos^2(\pi\theta)$ and $\left.\partial_\sigma \phi = -r e^{\phi}\right|_{\sigma=\pi}$ as the boundary condition. This boundary condition breaks the $SL(2,\mathbb{R})$ redundancy to $U(1)$.
The classical solution corresponding to these situations is:
\begin{equation}
e^{\phi} = -2\theta^2\frac{f'(u) f'(v)}{\sin\left(\theta \frac{f(u)-f(v)}{2}\right)^2},
\end{equation}
where one sets $\theta = n \in \mathbb{N}$ to find the ZZ${}_{1,n}$ brane again.\footnote{Strictly speaking, we should set $\theta \to i \theta$ to describe the genuine FZZT-branes, where the equation $r^2=\cosh^2(\pi\theta)$ becomes the standard FZZT relation; FZZT-branes correspond to hyperbolic orbits, whereas we described the elliptic orbits instead. Our choice of notation follows \cite{Dorn:2006ys}.}

Either of these alternative boundary conditions can be absorbed back into the action by rescaling $f \to \theta f$. The only effect is a change $\frac{2\pi}{\beta} \to \frac{2\pi}{\beta} \theta $ in \eqref{omega} and in the Hamiltonian in terms of $f$. After doing this, the field $f$ is again a circle diffeomorphism as before. \\

In all of these cases, we can make the link between the Liouville action in \eqref{pspi} and the geometric action of Alekseev and Shatashvili \cite{Alekseev:1988ce,Alekseev:1990mp} more explicit as follows. The $\pi_\phi\dot{\phi}$-term in \eqref{pspi} is precisely the canonical 1-form $\alpha$ integrated over time, with $\alpha = \int_{0}^{\pi} d\sigma\,  \pi \delta \phi, \quad \omega = d\alpha$. Given the symplectic 2-form $\omega$, and ignoring global issues, $\alpha$ is determined only up to an exact form $df$, which integrates to zero as we take periodic boundary conditions in time. Explicitly, and after doubling, the geometric action is given by:
\begin{equation}
\label{sgeom}
S_{\text{geom}} = \int d\tau \alpha = \int d\tau \int_{-\pi}^{\pi} d\sigma  \left[\frac{c}{48\pi}\frac{\dot{f}}{f'}\left(\frac{f'''}{f'}- 2 \left(\frac{f''}{f'}\right)^2 \right) - b_0 \dot{f}f'\right],
\end{equation}
with 
\begin{equation}
\alpha = \int_{-\pi}^{\pi} d\sigma  \left[\frac{c}{48\pi}\frac{df}{f'}\left(\frac{f'''}{f'}- 2 \left(\frac{f''}{f'}\right)^2 \right) - b_0 df \, f'\right],
\end{equation}
and $\omega = d\alpha$ given by equation \eqref{omega} as can be explicitly checked, and the coadjoint orbit parameter $b_0 = -\left(\frac{2\pi }{\beta} \frac{c}{24\pi} \theta\right)^2$ in terms of the FFZT brane parameter $\theta$. For the $\text{diff}(S^1)/U(1)$ orbit, one mods out by $F(\sigma,\tau) \to F(\sigma,\tau) + a(\tau)$, whereas for the $\text{diff}(S^1)/SL(2,\mathbb{R})$ orbit, one mods out by  $F(\sigma,\tau) \to \frac{\alpha(\tau)F(\sigma,\tau) + \beta(\tau)}{\gamma(\tau)F(\sigma,\tau) + \delta(\tau)}$, which indeed is what we find from the results of section \ref{sect:path} above. The function $F$ is as before the uniformizing coordinate, related to $f$ by $F = \tan \frac{\theta f}{2}$. \\
This result demonstrates the equivalence of Liouville between branes and the coadjoint orbit action (including the Hamiltonian term) for the different orbits depicted in Figure \ref{generalize}. At the level of the partition function, this also follows directly since both of these evaluate to the same Virasoro character. Indeed, computing characters of irreps of the algebra is precisely the goal of the coadjoint orbit construction:
\begin{equation}
\label{newpi}
\chi_h(T) = \int \left[\mathcal{D} f\right]e^{-\int_{0}^{T} \left(i\,  d^{-1}\omega + H d\tau\right)} = \int \left[\mathcal{D} f\right]e^{-S}, \qquad h \equiv h(b_0),
\end{equation}
with
\begin{equation}
\label{newgeoma}
S = \int d\tau \int_{-\pi}^{\pi} d\sigma  \left(i\left[\frac{c}{48\pi}\frac{\dot{f}}{f'}\left(\frac{f'''}{f'}- 2 \left(\frac{f''}{f'}\right)^2 \right) - b_0 \dot{f}f'\right] + \frac{c}{12\pi}\left\{\tan \frac{\theta f}{2},\sigma\right\}\right).
\end{equation}
This complicated expression can be transformed into the Floreanini-Jackiw path integral for a chiral boson, yielding indeed a single character \cite{Alekseev:1990mp}. On the other hand, it is known since the introduction of ZZ-branes in \cite{Zamolodchikov:2001ah} that the cylinder amplitude of Liouville between these ZZ-branes is computing the Virasoro vacuum character. As mentioned above, changing branes changes the character computed. \\
It is in any case reassuring to see this equality directly within the path integral. \\

The above procedure has the additional benefit that one now also has a dictionary between operator insertions in Liouville and operators in the Alekseev-Shatashvili geometric action theory \eqref{newpi},\eqref{newgeoma}. Explicitly, one has the correspondence
\begin{equation}
\label{ZZcorr}
e^{\ell \phi(\sigma,\tau)} \quad \leftrightarrow \quad \left(-2\theta^2\frac{f'(\sigma,\tau) f'(-\sigma,\tau)}{\sin\left(\frac{\theta}{2}(f(\sigma,\tau)-f(-\sigma,\tau))\right)^2}\right)^{\ell}.
\end{equation}
This correspondence is fully at the level of the 2d theories, and can be viewed as an interesting conclusion in its own right. \\

Finally taking the Schwarzian limit of interest, we need $|b_0| \to \infty$ such that
\begin{equation}
b_0 T^2 = -\left(\frac{2\pi }{\beta} C \theta\right)^2 \quad \Rightarrow \quad \theta = \frac{\beta}{2\pi}\frac{\sqrt{-b_0} T}{C}.
\end{equation}
As discussed in the main text, the above geometric action \eqref{sgeom} (the $p\dot{q}$-term in the Lagrangian) disappears in this limit, and only the Hamiltonian density (the Schwarzian derivative) remains.\footnote{The geometric action is identified in \cite{Oblak:2017ect} as a Berry phase associated to a closed path in the Virasoro group. Their holographic interpretation in AdS$_3$/CFT$_2$ in terms of precession of inertial frames agrees with their absence in the dimensionally reduced 2d JT gravity, dual to the Schwarzian.} As the Hamiltonian is itself the generator of a $U(1)$-symmetry, Stanford and Witten applied the Duistermaat-Heckman theorem to prove the one-loop exactness of the resulting 1d partition function \cite{Stanford:2017thb}. This one-loop exactness fails for correlation functions however and one has to resort to other methods, by using the correspondence \eqref{ZZcorr} in the Schwarzian limit, where the $\tau$-dependence drops out in \eqref{ZZcorr} and one recovers \eqref{corrrSchw} when $\theta=1$ to find the ZZ-ZZ system again. \\
When changing the branes, the resulting 1d theories are all pathological as thermal systems, except the ZZ-ZZ system that is studied here.

\section{Lagrangian description of matter sector}
\label{app:mattersch}
A general matter sector in the Poincar\'e upper half plane, is given by
\begin{equation}
S_m = \int df dz \, \mathcal{L}_m(q,\partial_f q),
\end{equation}
in terms of a canonical variable $q$. Consider now the coupling to the dynamical boundary as:\footnote{In path integral language, integrating the Jackiw-Teitelboim action \eqref{JTaction} over $\Phi^2$ fixed the AdS$_2$ metric; the Gibbons-Hawking boundary term then reduces to the Schwarzian action. \\
We chose here to perform the time reparametrization throughout the 2d bulk; the $z$-dependence of the fluctuating boundary is $\mathcal{O}(\epsilon)$ and can be ignored here.}
\begin{equation}
\label{tovary}
S = -\int dt \, \left\{f,t\right\} + \int dt dz \, f' \mathcal{L}_m(q,\partial_f q).
\end{equation}
The two sectors only interact through the dynamical time variable $f(t)$. As a sanity check, the matter equations of motion are given by
\begin{equation}
f' \frac{\partial \mathcal{L}_m}{\partial q} = \partial_t \left( f' \frac{\partial \mathcal{L}_m}{\partial \partial_f q} \frac{1}{f'}\right) \quad  \Rightarrow \quad \frac{\partial \mathcal{L}_m}{\partial q} = \partial_f \left(\frac{\partial \mathcal{L}_m}{\partial \partial_f q}\right),
\end{equation}
and are not influenced by the coupling to the dynamical boundary, as this is only a time reparametrization that indeed should not affect matter equations. \\
Varying \eqref{tovary} w.r.t. $f$ yields:
\begin{equation}
\delta S = \int dt \left[\frac{\left\{f,t\right\}'}{f'}\delta f - \int dx \left(-\mathcal{L}_m + \partial_f q \frac{\partial \mathcal{L}_m}{\partial \partial_f q}\right) \delta f' \right]=  \int dt \,\delta f \left[\frac{\left\{f,t\right\}'}{f'} + H'_m \right].
\end{equation}
The second term is found by writing $\partial_f q = \frac{q'}{f'}$ and using $\delta \frac{1}{f'} = - \frac{1}{{f'}^2} \delta f'$. This leads to 
\begin{equation}
\left\{f,t\right\}' + f'H'_m = 0,
\end{equation}
which matches the first derivative of eq. (3.16) of \cite{Engelsoy:2016xyb}.\footnote{Useful property:
\begin{equation}
\left(\frac{\left(\frac{f''}{f'}\right)'}{f'}\right)' = \frac{\left\{f,t\right\}'}{f'}.
\end{equation}}

\section{Partition function of a particle on a group manifold}
\label{app:partgroup}
Using the normalized eigenfunctions $\psi_j^{ab}(g) = \sqrt{\frac{d_j}{\text{vol }G}} (D_j)^{ab}(g)$ with $D_j(g)$ the representation matrices of the representation $j$, the Euclidean propagator from the point $g_0$ to $g_1$ with $e^\phi = g_1 (g_0)^{-1}$ is given by the formula \cite{Marinov:1979gm}:
\begin{equation}
K(g_0,g_1;t) = \sum_{j,a,b} \psi_j^{ab*}(g_0)\psi_j^{ab}(g_1) e^{-C_j t} = \frac{1}{\text{vol }G}\sum_{j} \text{dim j} \,\, \chi_j(\phi) e^{-C_j t}.
\end{equation}
with the character $\chi_j(\phi) = \text{Tr} \left[D_j(g_1) D_j^{\dagger}(g_0)\right]$. Setting $t=\beta$ and $\phi=0$, corresponding to the sum over all based loops on $G$, one finds $\chi_j(0)= \text{dim j}$ and
\begin{equation}
K(g_0,g_0;\beta) = \frac{1}{\text{vol }G}\sum_{j} (\text{dim j})^2 \, e^{-C_j \beta},
\end{equation}
which is indeed the path integral over $LG/G$ with the $(\text{vol }G)^{-1}$ factor coming from the right coset. This factor is absorbed into a contribution to the zero-temperature entropy $S_0$ and dismissed. The ordinary partition function of a particle on a group only contains the path integration over $LG$ and is indeed 
\begin{equation}
Z = \int dg_0 K(g_0,g_0;\beta) = \sum_{j} (\text{dim j})^2 \, e^{-C_j \beta}.
\end{equation}

\section{Some relevant formulas for $SU(2)_k$}
\label{app:tech}
The three-point function of the $SU(2)_k$ WZW model can be found in e.g. \cite{Gaberdiel:2007vu} as
\begin{equation}
\left\langle \Phi_{j_1,m_1,\bar{m}_1}(0)\Phi_{j_2,m_2,\bar{m}_2}(1)\Phi_{j_3,m_3,\bar{m}_3}(\infty)\right\rangle = \delta\left(\sum_i m_i\right) \delta\left(\sum_i\bar{m}_i\right) D_{j_1j_2}^{j_3}.
\end{equation}
In the special case that $m_2=\bar{m_2}=j_2$ and $m_3=\bar{m_3}=j_3$, one has explicitly:
\begin{align}
D_{j_1j_2}^{j_3} &= \frac{\Gamma(j_1+j_3-j_2+1)\Gamma(j_1+j_2-j_3+1)}{\Gamma(2j_1+1)}\sqrt{\frac{\gamma\left(\frac{1}{k+2}\right)}{\gamma\left(\frac{2j_1+1}{k+2}\right)\gamma\left(\frac{2j_2+1}{k+2}\right)\gamma\left(\frac{2j_3+1}{k+2}\right)}} \nonumber \\
&\times \frac{P(j_1+j_2+j_3+1)P(j_1+j_2-j_3)P(j_2+j_3-j_1)P(j_1+j_3-j_2)}{P(2j_1)P(2j_2)P(2j_3)}
\end{align}
with
\begin{equation}
\gamma(x) = \frac{\Gamma(x)}{\Gamma(1-x)}, \quad P(j) = \prod_{m=1}^{j}\gamma\left(\frac{m}{k+2}\right).
\end{equation}
In the large $k$ regime, we get
\begin{equation}
\gamma\left(\frac{\alpha}{k}\right) \,\, \to \,\, \frac{k}{\alpha}, \quad\quad P(j) \,\, \to \,\, \prod_{m=1}^{j}\frac{k}{m} = \frac{k}{\Gamma(j+1)},
\end{equation}
and hence
\begin{equation}
D_{j_1J}^{j_2} \, \to \, \sqrt{(2j_1+1)(2j_2+1)(2J+1)}\frac{\Gamma(2j_2+1)\Gamma(2J+1)}{\Gamma(j_1+j_2+J+2)\Gamma(j_2+J-j_1+1)}.
\end{equation}
On the other hand, we can write
\begin{align}
\left\langle \Phi_{j_1,m_1,\bar{m}_1}(0)\Phi_{J,M,\bar{M}}(1)\Phi_{j_2,m_2,\bar{m}_2}(\infty)\right\rangle &= \left\langle j_1,-m_1\right|\Phi_{J,M}\left|j_2,m_2\right\rangle
\nonumber \\
&= C_{J M,j_2m_2;j_1-m_1}^2 \tilde{A}_{j_1j_2J} \nonumber \\
&= C_{j_1 m_1,j_2m_2;J-M}^2 A_{j_1j_2J}
\end{align}
The second line uses the standard form of the Wigner-Eckart theorem by combining $j_2$ with $J$ into $j_1$. In the last equality we rearranged the Clebsch-Gordan coefficients using the symmetry of the $3j$-symbols; this reordering produces extra $j$-dependent factors that are absorbed into a new reduced matrix element $A_{j_1j_2J}$.
The Clebsch-Gordan coefficient for combining $j_1$ and $j_2$ into $J$ is:
\begin{align}
C_{j_1m_1,j_2,j_2;J-J} = &\sqrt{\frac{(2J+1)\Gamma(J+j_1-j_2+1)\Gamma(J+j_2-j_1+1)\Gamma(j_1+j_2-J+1)}{\Gamma(j_1+j_2+J+2)}} \nonumber \\
&\times (-)^{J-j_1-j_2}\sqrt{\Gamma(2J+1)\Gamma(2j_2+1)\Gamma(j_1+j_2-J+1)\Gamma(j_1+J-j_2+1)} \nonumber \\
&\times \sum_k \frac{(-)^k}{k!(j_1+j_2-J-k)!(j_1+J-j_2-k)!(-k)!k!(J+j_2-j_1+k)!},
\end{align}
with $m_1 = -m_2-m_3 = -j_2 + J$. It simplifies to 
\begin{align}
C_{j_1m_1,j_2,j_2;J-J} = \sqrt{\frac{(2J+1)\Gamma(2J+1)\Gamma(2j_2+1)}{\Gamma(j_1+j_2+J+2)\Gamma(J+j_2-j_1+1)}}(-)^{J-j_1-j_2}.
\end{align}
\end{appendix}

\end{document}